\documentclass[useAMS,usenatbib,fleqn]{mn2e}
\usepackage{times}
\usepackage{amsmath}
\usepackage{xfrac}
\usepackage{amssymb}
\usepackage{float}
\usepackage{stfloats}
\usepackage{balance}
\usepackage{graphicx}
\usepackage{caption}
\usepackage{subcaption}
\graphicspath{{./Figures/}}

\title[GAMA: Evolution of bias in the radio]{Galaxy and Mass Assembly (GAMA): The evolution of bias in the radio source population to $\bf{z\sim1.5}$}
\author[S. N. Lindsay et al.]
{\parbox{\textwidth}{S.~N.~Lindsay,$^1$\thanks{E-mail: s.lindsay2@herts.ac.uk} 
M.~J.~Jarvis,$^{2,3}$
M.~G.~Santos,$^{3,4}$
M.~J.~I.~Brown,$^5$ 
S.~M.~Croom,$^6$
S.~P.~Driver,$^{7,8}$ 
A.~M.~Hopkins,$^9$ 
J.~Liske,$^{10}$
J.~Loveday,$^{11}$
P.~Norberg$^{12}$ and
A.~S.~G.~Robotham$^{7,8}$} 
\vspace{0.4cm}\\ 
\parbox{\textwidth}{	
$^1$Centre for Astrophysics Research, Science \& Technology Research Institute, University of Hertfordshire, AL10 9AB, UK \\
$^2$Oxford Astrophysics, Department of Physics, Keble Road, Oxford, OX1 3RH, UK\\
$^3$Physics Department, University of the Western Cape, Bellville 7535, South Africa \\
$^4$CENTRA, Instituto Superior T\'ecnico, Universidade T\'ecnica de Lisboa, Av. Rovisco Pais 1, 1049-001 Lisboa, Portugal \\
$^5$School of Physics, Monash University, Clayton, Victoria 3800, Australia \\
$^6$ Sydney Institute for Astronomy, School of Physics A28, University of Sydney, NSW 2006, Australia \\
$^7$International Centre for Radio Astronomy Research (ICRAR), University of Western Australia, Crawley, WA 6009, Australia \\
$^8$Scottish Universities' Physics Alliance (SUPA), School of Physics and Astronomy, University of St Andrews, North Haugh, St Andrews, KY16 9SS, UK \\
$^9$Australian Astronomical Observatory, P.O.~Box 915, North Ryde, NSW 1670, Australia \\
$^{10}$European Southern Observatory, Karl-Schwarzschild-Str.~2, 85748, Garching, Germany \\
$^{11}$Astronomy Centre, University of Sussex, Falmer, Brighton BN1 9QH, UK \\
$^{12}$Institute for Computational Cosmology, Department of Physics, Durham University, Durham DH1 3LE, UK }}
\begin{document}

\date{\today}

\pagerange{\pageref{firstpage}--\pageref{lastpage}} \pubyear{2014}

\maketitle

\label{firstpage}

\begin{abstract}
We present a large-scale clustering analysis of radio galaxies in the Very Large Array (VLA) Faint Images of the Radio Sky at Twenty-cm (FIRST) survey over the Galaxy And Mass Assembly (GAMA) survey area, limited to $S_{1.4 \textrm{ GHz}}$ \textgreater 1~mJy with spectroscopic and photometric redshift limits up to $r < 19.8$ and $r <22$~mag, respectively. For the GAMA spectroscopic matches, we present the redshift-space and projected correlation functions, the latter of which yielding a correlation length $r_0 \sim 8.2$ $h^{-1}$Mpc and linear bias of $\sim1.9$ at $z\sim0.34$. Furthermore, we use the angular two-point correlation function $w(\theta)$ to determine spatial clustering properties at higher redshifts. We find $r_0$ to increase from $\sim6$ to $\sim 14$ $h^{-1}$Mpc between $z=0.3$ and $z=1.55$, with the corresponding bias increasing from $\sim 2$ to $\sim 10$ over the same range. Our results are consistent with the bias prescription implemented in the SKADS simulations at low redshift, but exceed these predictions at $z>1$. This is indicative of an increasing (rather than fixed) halo mass and/or AGN fraction at higher redshifts or a larger typical halo mass for the more abundant FR\textsc{I} sources.

 \end{abstract}

\begin{keywords}
surveys -- galaxies: active -- cosmology: large-scale structure of Universe -- radio continuum: galaxies
\end{keywords}

\section{Introduction}

In recent decades, measurements of the cosmic microwave background radiation (CMB; e.g. \citealt{komatsu11}) have proven the very early Universe to be remarkably isotropic with tiny under- and overdensities that have grown to form the vast and intricate structures we see today. Galaxies and clusters observed today are far removed from their almost homogeneous beginnings and we require large numbers of them to piece together a statistical picture on cosmological scales. Clustering measures on large scales can be used to investigate not only the relationships between galaxy populations found by various techniques probing different epochs and masses, but also broader cosmological phenomena such as baryon acoustic oscillations (BAO; e.g. \citealt{eisenstein05, percival10, blake11}), cosmic magnification (e.g. \citealt{scranton05, wang11}), or the integrated Sachs-Wolfe (ISW) effect (e.g. \citealt{mcewen07, giannantonio08, raccanelli08}). 

Cosmological applications require information about the gravitating mass distribution in the Universe, which in a $\Lambda$CDM cosmology is strongly tied to the dark matter distribution. Direct observations tell us only about the baryonic matter, from which we must infer the dark matter distribution. Various tools exist for measuring the clustering signal of an observed source catalogue, such as nearest neighbour measures (e.g. \citealt{bahcall83}), counts-in-cells (e.g. \citealt{magliocchetti99,blake02b,yang11}), correlation functions (e.g. \citealt{groth77,bahcall83,blake02a,blake02b,croom05}) and power spectra (e.g. \citealt{cole05,percival07,komatsu11}). Due to its relative simplicity to calculate, and relation to its Fourier transform (the power spectrum), the two-point spatial correlation function has become a standard in quantifying cosmological structure. A means by which we can quantify the extent to which the observable and dark matter are tied using the correlation function is through the bias parameter $b(z)$. This relates the spatial correlation function (see Section \ref{spatial}) of an observed galaxy population to that of the underlying dark matter. The bias quantifies the difference in the clustering of the dark matter haloes acting solely under gravity and of galaxies inhabiting those haloes with other effects making their structure more or less diffuse. This has a heavy dependence on the galaxy masses and the epoch under consideration (e.g. \citealt{seljak04}).

Local galaxies with masses comparable to the Milky Way are a relatively unbiased tracer of mass, with a present day correlation length (the clustering scale at which the correlation function falls below unity; see Section \ref{spatial}) of $r_0 = 5.4$ $h^{-1}$Mpc found in the early CfA redshift survey \citep{davis83}. While the power law slope is consistently found to be $\sim 2$ in most studies, large variations in $r_0$ are found depending on the populations and respective epochs being observed, from $\sim 5$ $h^{-1}$Mpc for local galaxies to $\sim 25$ $h^{-1}$Mpc for Abell clusters (e.g. \citealt{bahcall83}). At the lower end of the clustering scale, \citet{saunders92} find $r_0 \sim 4$ $h^{-1}$Mpc for IRAS starburst galaxies and using the 2dF QSO Redshift Survey \citet{croom01} show a scale roughly constant with redshift at $\sim 5$ $h^{-1}$Mpc for quasars at $0 \lesssim z \lesssim 2.5$ (see also \citealt{croom05,ross09,ivashchenko10}). Likewise, \citet{kovac07}  find that $z \sim 4.5$ Ly-$\alpha$ emitters exhibit a relatively short clustering length of $4.6$ $h^{-1}$Mpc, consistent with their being progenitors of Milky Way type local galaxies (see \citealt{nilsson09} and references therein), and note the similarity with Lyman break galaxies (LBGs) at $z \sim 3.8$ and $z \sim 4.9$, suggesting that the two populations reside in the same host halos but with a relatively low duty cycle. Local $L \gtrsim L_{\ast}$ ellipticals, however, are a highly clustered population with various authors finding $r_0 \sim 7$--$12$ $h^{-1}$Mpc (e.g. \citealt{guzzo97,willmer98,norberg02}). Similarly high clustering lengths are found for extremely red objects (EROs) at $z \sim 1$ (e.g. \citealt{daddi01,mccarthy01,roche02}) and distant red galaxies (DRGs) at $z \sim$1--2 \citep{grazian06,foucaud07}, both posited as progenitors of the local bright ellipticals. Indeed, similar results for radio galaxies (e.g. \citealt{cress96,overzier03}) give weight to the suggestion by \citet{willott01} that EROs and radio galaxies are identical and seen at different evolutionary stages, based on their findings of ERO-like hosts of radio galaxies in the 7C Redshift Survey.

Radio surveys are ideal for the purpose of carrying out such large-scale statistical measurements for a number of reasons. From a logistical point of view, radio wavelengths are useful because they occupy a uniquely broad atmospheric window in the electromagnetic spectrum, not being significantly absorbed by Earth's atmosphere. This allows observations to be carried out from terrestrial telescopes, reducing costs compared with launching and operating space telescopes in orbit. In a wide radio survey with relatively shallow flux density limit ($S_{1.4} \gtrsim 1$ mJy) extragalactic sources dominate in the form of synchrotron radiation in active galaxies with supermassive black holes at their centre. By observing such powerful sources, and with the long radio wavelengths being immune to absorption by dust, the ISM and our own atmosphere, ground based telescopes are able to observe the whole sky relatively unobscured and to very high redshifts. 

Of particular significance for the science discussed here is that these radio-loud active galactic nuclei (AGN) predominantly reside in massive elliptical galaxies (e.g. \citealt{jarvis01,willott03,mclure04,herbert11}) and, as such, provide good tracers of large-scale clustering and the underlying dark matter distribution. Radio surveys have been exploited with this aim for some time, such as the 4.85 GHz 87GB \citep{kooiman95} and Parkes-MIT-NRAO \citep{loan97} surveys, the 325 MHz Westerbork Northern Sky Survey (WENSS; \citealt{rengelink98}) and 1.4 GHz VLA surveys, Faint Images of the Radio Sky at Twenty-cm (FIRST; \citealt{becker95}) and the NRAO VLA Sky Survey (NVSS; \citealt{condon98}) (e.g. \citealt{cress96,blake02a,overzier03, best05, fine11}), but their potential uses are limited without additional data at other wavelengths. While the large redshift range of radio surveys can be advantageous over more targeted optical surveys, this can also wash out the clustering signal due to the superposition of the influence of differently behaving subpopulations (e.g \citealt{wilman08}). This effect can be alleviated by the use of complementary optical surveys to aid the separation of radio sources by simple redshift binning or other methods. 
	
As technology advances, observations at radio wavelengths will be able to constrain cosmological parameters with more precision than ever \citep{raccanelli12, camera12}, as well as provide the data for much more galactic and extragalactic science \citep{norris11}. Moving forward, the large numbers of sources observed in various overlapping surveys will make multi-wavelength studies not only more common, but essential to reliably identify objects by their emission across the entire spectrum and delineate the clustering properties of various subpopulations of any given individual survey.
	
It is with this goal in mind that in this paper we combine the available radio data from the FIRST survey, and optical/near-infrared photometry and spectra from Galaxy and Mass Assembly (GAMA; \citealt{driver11}), Sloan Digital Sky Survey (SDSS; \citealt{york00}) and UKIRT Infrared Deep Sky Survey (UKIDSS; \citealt{lawrence07}) to measure the large scale clustering properties of the $S_{1.4} > 1$ mJy radio emitting galaxies. We choose to use FIRST over NVSS for its marginally greater depth and considerably better resolution which is essential for meaningful cross-identification with the optical surveys. Furthermore, we employ a parent redshift distribution from the simulated radio catalogue of the SKA Design Study (SKADS; \citealt{wilman08}) which has the express intention of acting as a test bed to inform the use of the forthcoming Square Kilometre Array.
	
The outline of this paper is as follows: Section \ref{data} describes the surveys and catalogues used, while Sections \ref{angular} and \ref{Limber} detail the angular correlation function measurements and their deprojection to infer spatial clustering properties. Section \ref{spatial} describes direct measurements of the spatial clustering properties through the spatial correlation function and projected correlation function and Section \ref{massbias} describes the linear bias found from each of these methods out to $z\sim1.5$. In Section \ref{conclusions} we summarise our results and conclusions. 
	
The cosmological model used throughout this paper is the flat, $\Lambda$CDM concordance cosmology where $\Omega_m = 0.3$, $\Omega_\Lambda = 0.7$ and $\sigma_8 = 0.8$. All distances are kept in units of $h^{-1}$Mpc where $H_0 = 100 h$ km s$^{-1}$Mpc$^{-1}$ and $h$ is not explicitly assumed.

\section{Data}\label{data}

To infer the spatial clustering parameters of the radio galaxy population, redshifts of the sources (or their redshift distribution, at least) are required. We use spectroscopic redshifts from the GAMA survey, supplemented by photometric redshifts calculated using optical and near infrared photometry from SDSS and UKIDSS (described in \citealt{smith11}). These redshift catalogues are cross-matched with radio sources from the FIRST survey to assign optical counterparts and redshifts to each radio source. 

\subsection{Radio Surveys}\label{radio}

The FIRST survey \citep{becker95} was carried out at an observing frequency of 1.4 GHz with the Very Large Array (VLA) in B configuration. The most recent catalogue\footnote{\it{http://sundog.stsci.edu/first/catalogs/readme\_12feb16.html}} contains 946,464 sources covering over 10,000 square degrees ($\sim$8,500 in the northern galactic cap and $\sim$1,500 in the south) with an angular resolution of 5.4\arcsec (FWHM) to a completeness of 95 per cent at $S_{1.4\rmn{GHz}} >$ 2 mJy. A number of sidelobes are spuriously counted in the raw catalogue and an oblique decision-tree program developed by the FIRST survey team \citep{white97} finds probabilities of each catalogue entry representing extended activity from a nearby bright source. Those entries with a sidelobe probability of $>$0.1 have been excluded for the purposes of this analysis ($<$20 per cent of the catalogue), leaving 723,934 sources above 1 mJy.

\begin{figure}
\centering
\includegraphics[width=0.48\textwidth]{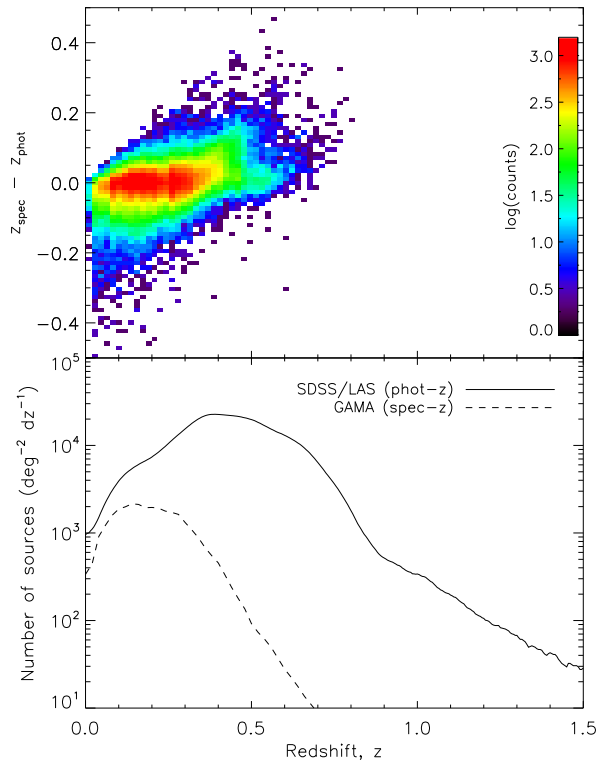}
\caption{\textit{Top:} Comparison of SDSS/LAS photometric redshift estimates with GAMA spectroscopic ($Q \geq 3$) redshifts. \textit{Bottom:} Redshift distributions of the full photometric and spectroscopic catalogues (galaxies only).}
\label{GAMAz}
\end{figure}

\subsection[]{Optical and Near-Infrared Surveys}\label{optical}
	
The Galaxy And Mass Assembly (GAMA; \citealt{driver11}) survey has been in operation since 2008, using the 3.9m Anglo-Australian Telescope (AAT) and the AAOmega spectrograph to build a $>$98 per cent complete redshift catalogue of $\sim$140,000 galaxies to a depth of $r <$ 19.4 or 19.8. Two gratings with central wavelengths of 4800{\AA} and 7250{\AA}  are used, giving continuous wavelength coverage of the range 3720--8850{\AA} with resolution $\sim 3.5${\AA}  (in the blue channel) and $\sim$ 5.5{\AA}  (in the red channel). Spectroscopic redshifts are found in real time using the \textsc{runz} code developed for the 2dFGRS \citep{colless01} and given a quality flag $Q$ of 0--4 where $Q \geq 3$ is a probable or certain redshift worthy of publication \citep{hopkins13}. Corresponding photometry is also available in UKIDSS and SDSS bands \citep{hill11}. 
	
GAMA is divided over three 12$\times$4 degree equatorial regions centred at $\alpha =$ 135\degr (9h), 180\degr (12h) and 217.5\degr (15h), covering 144 square degrees. These were chosen to be large enough in area to fully sample $\sim$100 $h^{-1}$Mpc structures at $z \sim 0.2$, and for their overlap with existing surveys. The data used for this paper come from the first 3 years of observations (GAMA Phase I; \citealt{driver11}).
	
The overlap of the GAMA survey area with other surveys allows for the collection of sufficient photometry to be able to generate a large contribution to the redshift catalogue using photometric redshifts. The Sloan Digital Sky Survey (SDSS) has $\sim$10,000 square degrees of sky coverage in 5 bands ({\it ugriz}), with the southernmost stripes covering the 3 GAMA fields. The UKIRT Infrared Deep Sky Survey (UKIDSS) uses the 3.8m UKIRT consisting of several subsurveys, including the Large Area Survey (LAS). UKIDSS LAS complements the SDSS, covering over 2,000 square degrees in the {\it YJHK} bands within the SDSS regions and overlapping with GAMA. We therefore have photometry in up to 9 optical and infrared bands with which to obtain photometric redshifts for those galaxies without spectroscopy (further details are given in \citealt{smith11}). For those sources detected in the SDSS imaging data but without GAMA spectroscopic redshifts, we impose a magnitude limit of $r <$ 22, i.e. within the limit of the SDSS imaging data. Figure \ref{GAMAz} shows a comparison between spectroscopic and photometric redshifts, where we find an rms difference of 0.07, and their distributions	.
	
\begin{table}
\begin{center}
\caption{Spatial boundaries and surface densities ($\overline{\sigma}$) for the three GAMA/SDSS/LAS fields used (comprising both spectroscopic and photometric redshifts). Due to incomplete coverage in one corner of the 9h field, the area is $\sim 0.4$ sq. deg. smaller in area than the full rectangular 12h and 15h fields. All subsequent analysis accounts for this.}
\begin{tabular}{l c c c r} \hline	
Field & RA range (\degr) & Dec range (\degr) & Optical sources & $\overline{\sigma}$ (deg$^{-2}$) \\ \hline 
9h & [128.0, 142.0] & [-2.0, 3.0]  & 633,229 & 9,101 \\ 
12h & [172.5, 186.5]  & [-2.9, 2.1]  & 651,148 & 9,303 \\ 
15h & [210.5, 224.5]  & [-2.0, 3.0]  & 653,417 & 9,336 \\ \hline
\end{tabular}
\label{GAMAfield}
\end{center}	
\end{table}

\begin{table}
\begin{center}
\caption{Areas and surface densities of $S_{1.4}>1$ mJy radio sources in the FIRST catalogues, the GAMA counterparts to the radio sources and also for the SKADS simulated data set. Note that the low-density of sources in the whole (post-collapse) FIRST catalogue compared to the prediction from SKADS may be explained by the incompleteness at $<2$~mJy (Sections~\ref{completeness} and~\ref{variance}) and the effects of resolution bias (Section~\ref{redshifts})}
\begin{tabular}{l c c r} \hline
Catalogue & No. of sources & Area (deg$^2$) & $\overline{\sigma}$ (deg$^{-2}$) \\ \hline 
\textsc{FIRST} & 585,473 & $\sim$10,000 & $\sim$58.5 \\ 
\textsc{GAMA} & 3,886 & 210 & 18.5 \\ 
\textsc{SKADS} & 32,061 & 398 &  80.6 \\ 	\hline
\end{tabular}
\label{catalogues}
\end{center}
\end{table}

\begin{figure}
\centering
\includegraphics[width=0.48\textwidth]{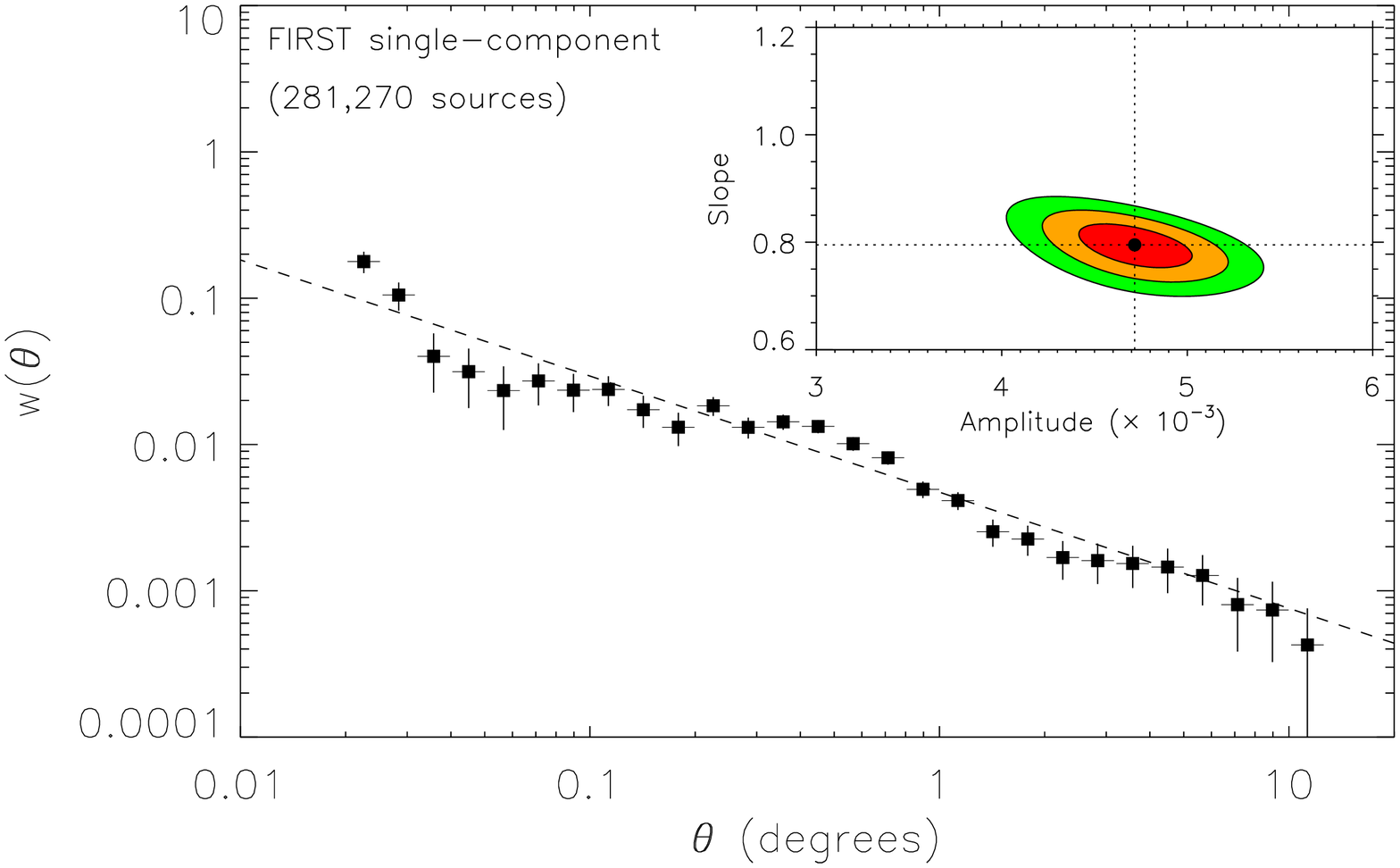}
\includegraphics[width=0.48\textwidth]{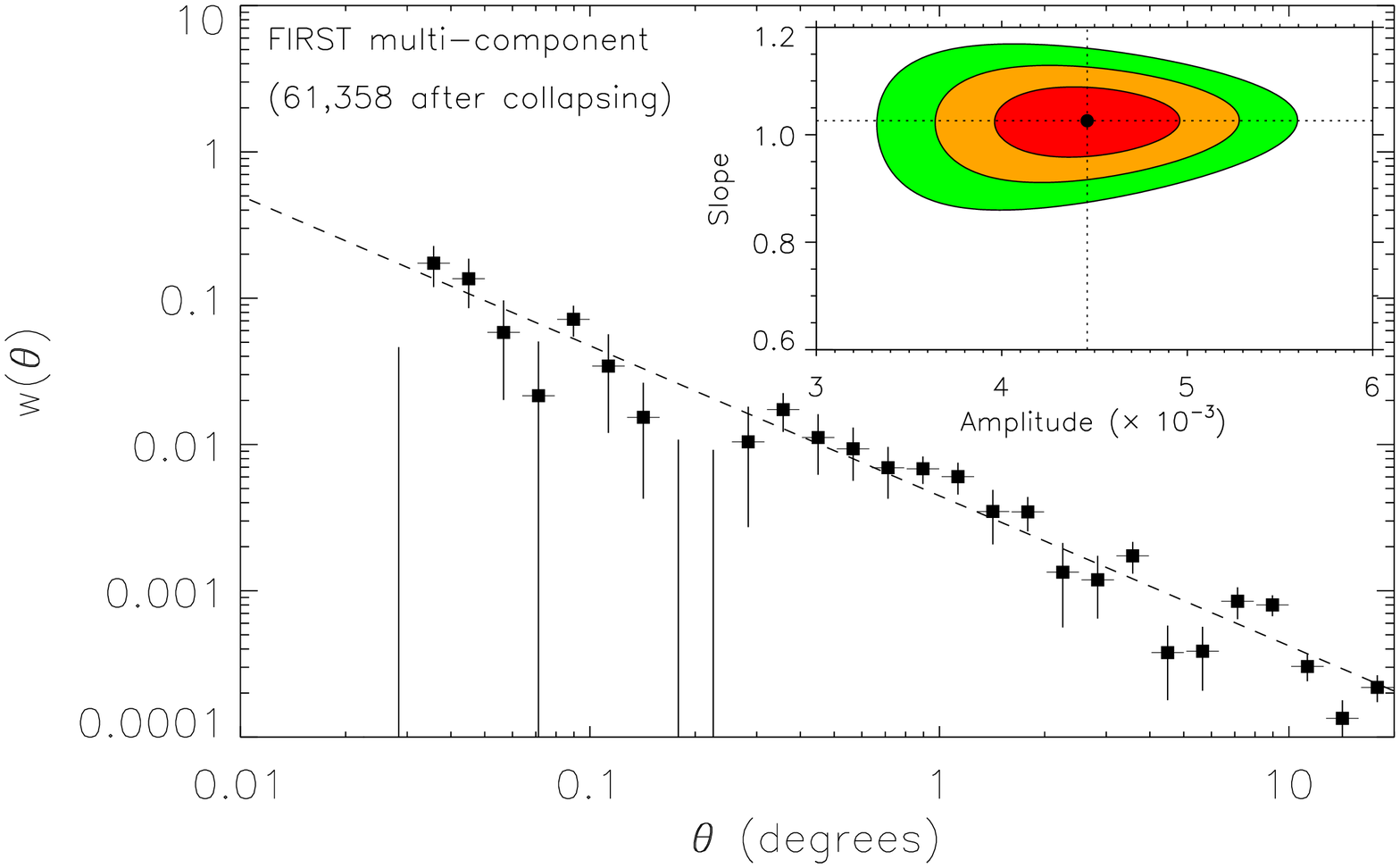}
\caption{Angular correlation functions for a 115\degr $\times$ 64\degr area of the FIRST survey at 1 mJy, with 68, 90 and 95 per cent contours for the power-law parameter fits. The top panel corresponds to those FIRST sources which have no neighbours within the 72\arcsec collapsing radius adopted. The bottom panel corresponds to the remaining (assumed) extended sources post-collapsing.}
\label{FIRSTw}
\end{figure}		
	
\subsection{Collapsing of Multi-Component Radio Sources}\label{collapsing}	

In order to address the inevitable issue of extended radio sources resulting in multiple detections for one host galaxy, of which perhaps none corresponds to the core itself (and therefore any associated optical source), we have followed \citet{cress96} in applying a collapsing radius of 72\arcsec (0.02\degr) to the FIRST catalogue. Any FIRST sources within this radius of one another are grouped and combined to form a single entry positioned at the flux-weighted average coordinates of the group and attributed with their summed flux density. This precludes us probing the correlation function to angular scales smaller than 72\arcsec, but ensures that only extremely extended sources will be mistakenly treated as independent galaxies. The procedure does, however, suffer from introducing far greater positional uncertainty to the assumed core position given the inherent variation in morphology in the observed objects. Asymmetric multiple-component galaxies or independent galaxies with small angular separations by chance, in particular, will interfere with the reliability of the collapsed catalogue. From the original FIRST catalogue, 227,606 sources are collapsed in 98,225 distinct groups, leaving a $>$1 mJy catalogue of 585,473 sources.
	
Returning to the option of supplementing FIRST with NVSS data to better deal with extended sources, when matching collapsed FIRST sources or NVSS sources directly to the optical catalogues, we find no evidence of a significant difference between the results. 
	
	
Figure \ref{FIRSTw} shows the angular correlation function (discussed in Section \ref{angular}) of a large area (115\degr $\times$ 64\degr) of the FIRST survey, separately for single sources and collapsed extended sources. The single sources demonstrate a well constrained power law form while the collapsed sources give more erratic results, beyond the simple Poisson errors due to their smaller numbers. However, the results are still broadly compatible and the signal of the combined, collapsed catalogue is dominated by the well-behaved single source power law. The two samples would not be expected to have the same $w(\theta)$, however, as the collapsed sources are dominated by extended FR\textsc{I} and FR\textsc{II} AGN, in contrast to the largely compact single-component sources, which are a combination of compact AGN and low-redshift star-forming galaxies.
	
\begin{figure}
\centering
\includegraphics[width=0.49\textwidth]{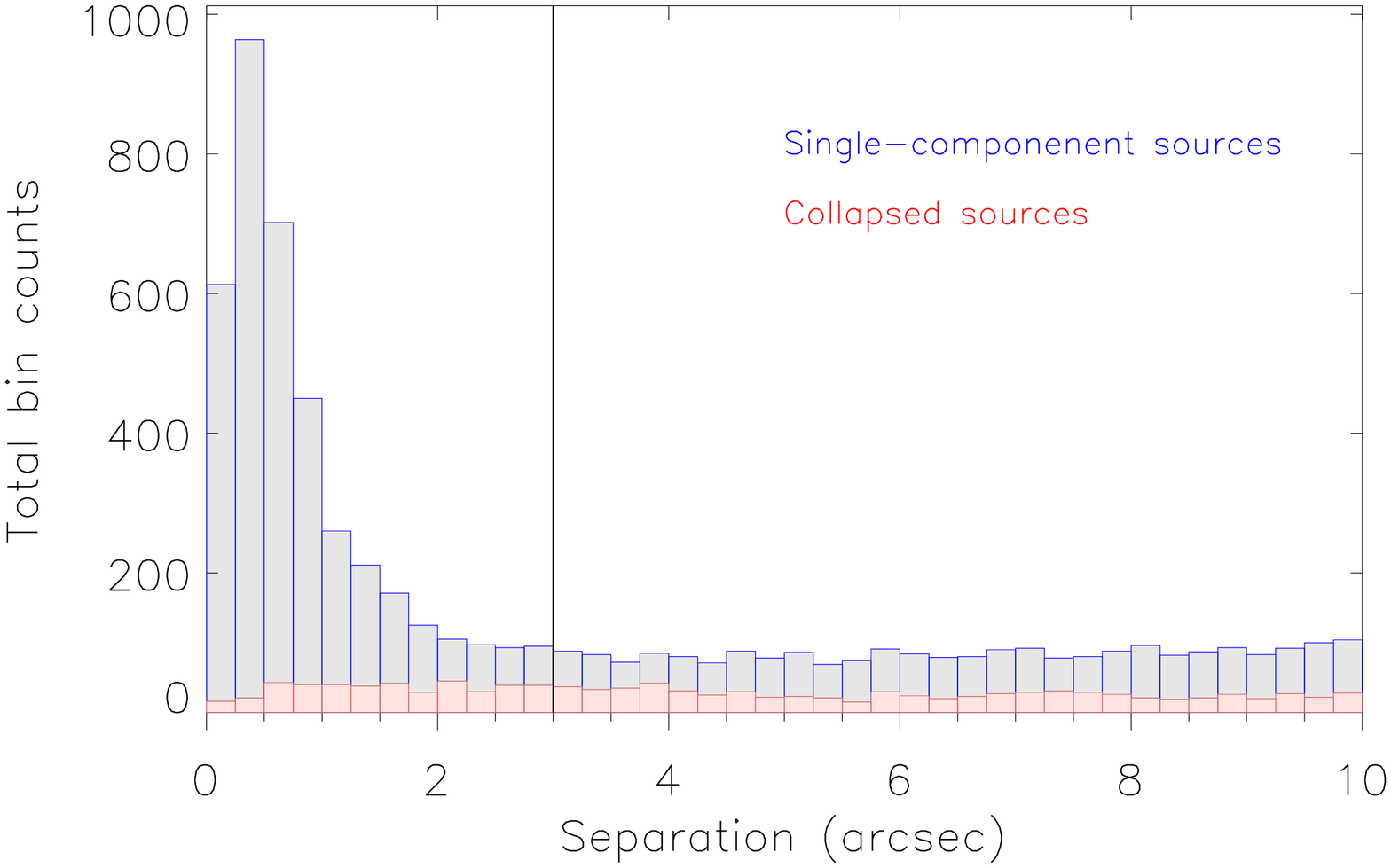}
\includegraphics[width=0.49\textwidth]{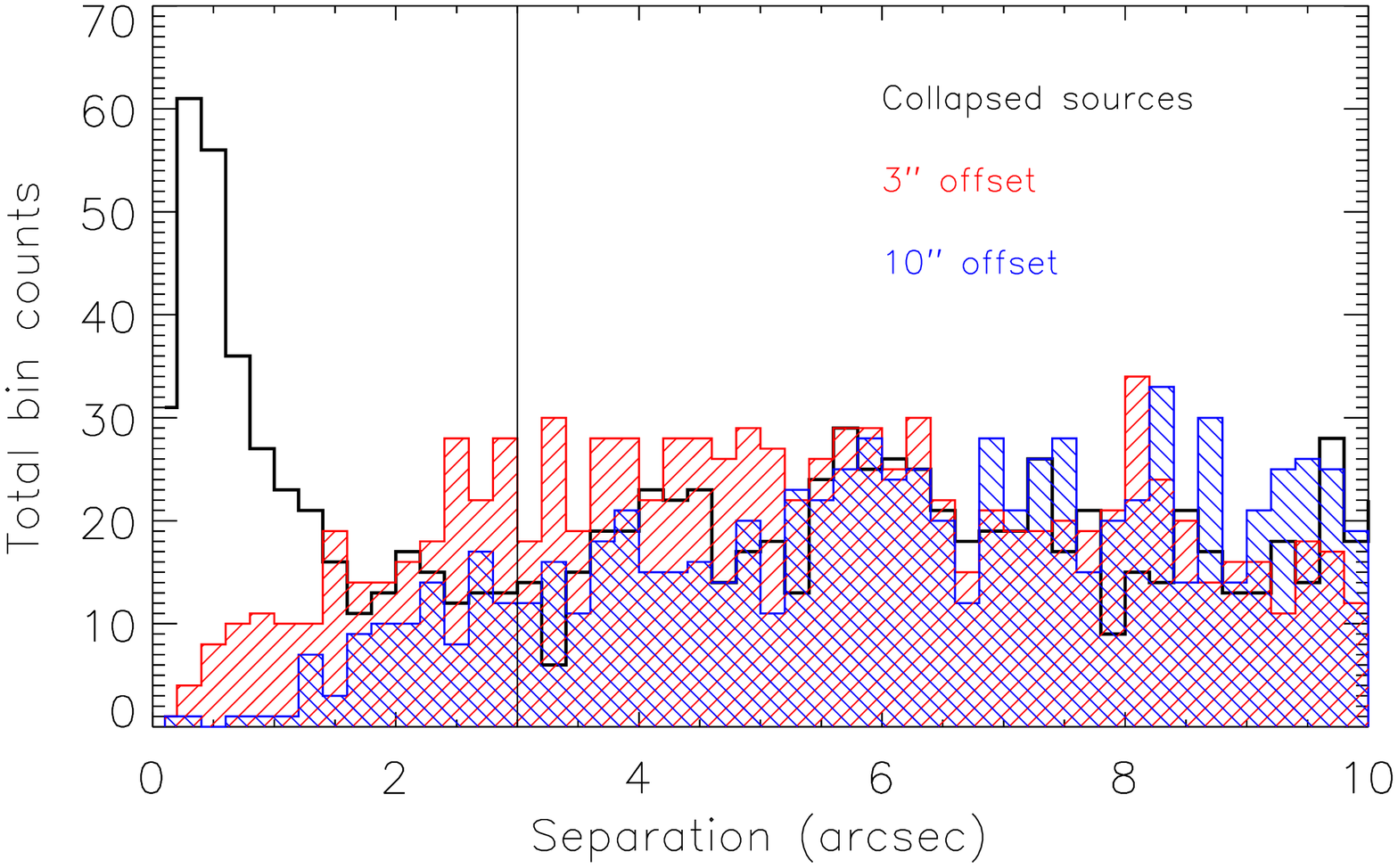}
\caption{\textit{Top}: Distribution of offsets between FIRST sources and the nearest GAMA/SDSS source, with single-component sources (\textit{blue}) placed on top of the collapsed sources (\textit{red}). The vertical solid line indicates the proposed separation cut-off of 3\arcsec\   to define a good radio-optical cross-match, below which counts rise above the background level. Collapsed sources account for $\sim 10$ per cent of matches within the cut-off. \textit{Bottom}: Distribution of separations for the collapsed radio sources (\textit{black}). The filled histograms show the result of assigning a random Gaussian offset of $\sigma=$3\arcsec (\textit{red}) and 10\arcsec (\textit{blue}) before the cross-matching showing a strong decline in matches within 2\arcsec.}
\label{sephist}		
\end{figure}
		
\subsection{Optical Identification of FIRST Sources}\label{matching}
	
An optical catalogue somewhat larger than the GAMA survey area is used for the purpose of cross-matching with the radio sources (see Table \ref{GAMAfield}), making use of photometric redshifts determined using up to 9 SDSS/UKIDSS-LAS bands ({\it ugrizYJHK}). This allows for greater source counts and improved statistics for the analysis, as well as expanding to a region completely containing the GAMA fields as well as other surveys like \textit{Herschel}-ATLAS \citep{eales10}. The expanded region comprises three 14 x 5 degree fields expanding from the three 12 x 4 degree GAMA fields.
	
Of the $\sim$ 6 million optical/IR sources in this area with at least a photometric redshift, approximately half are removed on the basis of the star-galaxy separation technique by \citet{baldry10}, leaving only the likely extragalactic sources to be matched to the radio. This separation is based on $J - K$ and $g - i$ colours rather than simply removing point-like sources and therefore fewer quasars are mistakenly discarded, provided they lie away from the stellar locus. As a second measure, a minimum redshift criterion of $z>0.002$ was added to filter out the nearby objects (some with significantly negative redshifts) which are assumed to be stellar in origin or to correspond to highly extended extragalactic sources, although we note that this may also filter out a small number of real low-redshift extragalactic radio sources. 

For the purpose of radio-optical cross-matching, given the positional accuracies of the catalogues ($<$1\arcsec\ for FIRST, $\sim$ 0.1\arcsec\ for SDSS) a simple nearest-neighbour match can be reliable. \citet{sullivan04} find this method to produce very similar catalogues to the likelihood ratio protocol of \citet{sutherland92} in their work. The likelihood ratio technique is often used to identify radio sources \citep{gonzalez05,ciliegi05,afonso06,mcalpine12} but here we instead use the simpler method tested by \citet{sullivan04} and adopted by e.g. \citet{elbouchefry07}. The 13,346 radio sources within the three fields were all found to have optical counterparts within 1 arcminute. By inspection of the separation distribution shown in Figure \ref{sephist}, any separations of more than a few arcseconds are likely to be random. A compromise must be reached between including the greatest number of our radio sample and ensuring reliable optical cross-IDs. The expected contamination of the remaining objects by chance proximity of optical sources can be expressed through the simple equation:
\begin{equation}
P_c = \pi \times r_s^2 \times \sigma_\textrm{opt.},
\end{equation}
where $r_s$ is the search radius cut-off and $\sigma_\textrm{opt.} = 9.23 \times 10^3$ deg$^{-2}$, the surface density of the optical/IR galaxy catalogue. This gives a 0.9 per cent contamination rate for a 2\arcsec\ cut-off and 2.0 per cent for a 3\arcsec\ cut-off. 
We place our separation limit at 3\arcsec, below which the distribution in Figure~\ref{sephist} becomes visibly more dense. This leaves 3,886 (29 per cent) of the original FIRST radio sources, of which 422 correspond to collapsed multi-component sources, an estimated 78 are spurious and 1424 (42 per cent) have good quality GAMA spectroscopic redshifts. 	

While positional coincidence is appropriate for defining positive identifications of single isolated radio sources, it is not necessarily so for the collapsed multiple sources, as this introduces uncertainty into the assumed position of the optical core of the host galaxy. These sources are in the minority, but to assess the reliability of their matches, we repeat the cross-matching process with the radio positions randomly displaced by 3\arcsec and 10\arcsec Gaussian distributions. Figure \ref{sephist} (bottom) shows the resulting separation histograms. While all three iterations of the procedure agree above the 3\arcsec matching cut-off, the significant peak at low separations is completely lost after shifting the radio positions. There will almost certainly be a minority of collapsed sources for which we do not accurately find a radio core position, but this test shows that for those where the collapsing is successful, reliable optical matches are being found.

\begin{figure}
\includegraphics[width=0.48\textwidth]{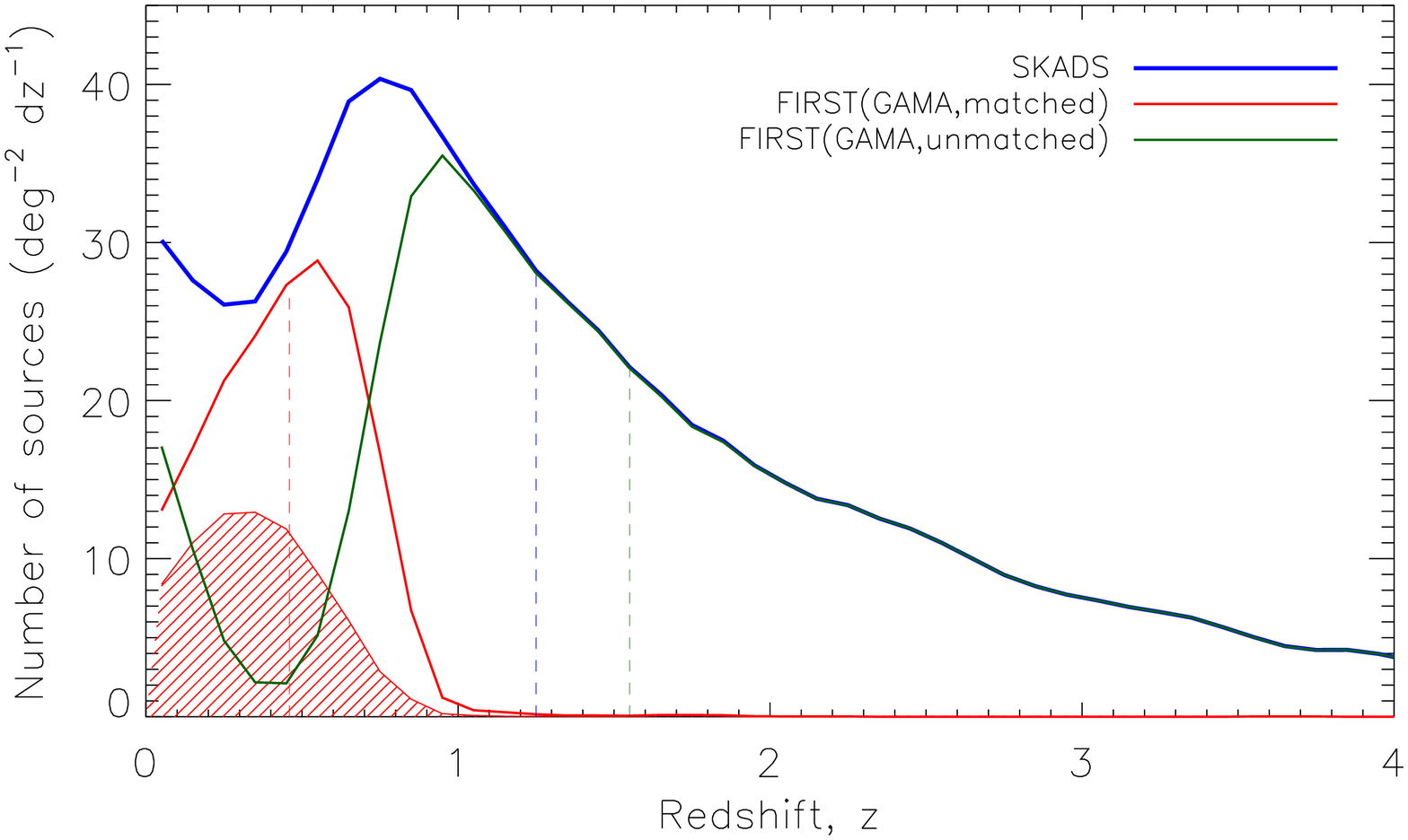}
\caption{Redshift distributions of the 1 mJy \textsc{SKADS} (\textit{blue}) and \textsc{GAMA} catalogues (\textit{red}), where the SKADS $N(z)$ is assumed to be the redshift distribution of FIRST sources. The distribution of the radio sources not identified in GAMA (\textit{green}) is inferred assuming a SKADS-like parent distribution and subtracting the \textsc{GAMA} distribution. Dotted lines mark the median redshifts for each set of objects and the filled distribution describe the GAMA spectroscopic redshifts (i.e. removing SDSS/LAS photometric redshifts).}
\label{zdists}
\end{figure}	

\subsection{Redshift Distributions}\label{redshifts}

In order to study the spatial clustering properties of our galaxy samples, we require knowledge of their redshifts, either individually or at least their distribution. We have spectroscopic or photometric redshifts for those FIRST radio sources with optical identifications, but do not have direct information for the optically unidentified radio sources. To estimate the complete redshift distribution as a function of radio flux density limit, we use the SKADS simulation which is based on a range of observed luminosity functions (see \citealt{wilman08} for full details).

Figure \ref{zdists} shows the redshift distribution of the SKADS catalogue at 1 mJy and the cross-matched subset of FIRST. By making the assumption that FIRST sources (the wider catalogue as well as just those within the GAMA fields) should have a similar distribution to that from SKADS, we may compare directly the clustering of the real and simulated catalogues. Furthermore, as shown in Figure \ref{zdists}, a third distribution can be inferred by subtracting the cross-matched distribution from the SKADS ``parent'' distribution (following similar work by \citealt{passmoor13}). This describes the remaining unmatched radio sources with no direct redshift measurements. 

It is clear that the distribution of unmatched sources shows an up-turn towards low-redshift ($z< 0.2$), where there appears to be a large fraction of sources in the SKADS simulation that are either not detected in FIRST or are in FIRST but do not have a counterpart in GAMA. The latter explanation of this is unlikely, as we would expect that the majority of relatively bright radio sources at $z< 0.2$ to have a reasonably bright optical counterpart. We therefore suggest that these sources, which are dominated by star-forming galaxies in the SKADS simulation and are predicted to be there based on low-redshift far-infrared luminosity functions (see \citealt{wilman10}), are likely to be resolved out by the VLA observations in the B-Array configuration used for the FIRST survey (see also \citealt{jarvis10}).

\cite{simpson12} also noted a similar deficit of low redshift sources compared with SKADS predictions in their radio survey of the Subaru/XMM-Newton Deep field, and attributed this to the effects of resolution bias.  This resolution bias refers to the number of faint resolved sources missing from a peak-flux-density-limited survey because their extended emission is not detected by the radio interferometer. \cite{bondi03} performed detailed simulations to determine the effect of this bias on their deep VLA catalogue and their estimates suggest that around 25 per cent of sources could be missing due to this resolution bias, and that these would be concentrated at low redshift. Any low-redshift star-forming galaxies that have radio flux-density detectable in FIRST are unlikely to fall below the SDSS depth ($r <$ 22), although low-surface brightness optical incompleteness may play a role if the radio data were deeper. 

Thus we suggest that the discrepancy between the simulated redshift distribution and the observed redshift distribution at low redshift is consistent with being due to resolution bias. We also note that some of the sources in our cross-matched sample could also be removed by our removal of sources with photometric redshifts $z<0.002$.


\subsection{Completeness}\label{completeness}

While the FIRST radio catalogue contains a large number of  $\sim 1$\,mJy sources, it is incomplete below 2--3 mJy. Discarding these fainter radio sources would significantly affect the size of our samples and the statistical significance of our results, but the effects of the incompleteness at the mJy level must be corrected for. 
To account for the noise variations across the survey area we use an rms noise map of the FIRST survey and their source detection criterion of $S - 0.25\textrm{(mJy)} > 5\times \textrm{rms}$ to determine whether a radio source randomly drawn from the SKADS simulated catalogues would be detected in the FIRST data over the GAMA fields. We use only those sources from SKADS with $S_{1.4} > 0.5$\,mJy, and if the SKADS radio source fulfills the FIRST survey detection criteria then that source is retained, whereas if the source falls below the detection threshold then the source is omitted. This process not only provides an estimate of the completeness of the survey but also allows us to compile a catalogue of sources, with both flux density and redshift information, that is subsequently used as our random source catalogue for calculating the angular and spatial correlation function (see Secs.~\ref{angular}, \ref{Limber} and \ref{spatial}). This ensures that we fully account for any noise variation and incompleteness across the survey area. The model redshift distribution of radio sources in our survey regions is thus modified according to these noise variations and it is this modified redshift distribution which is used throughout in determining the spatial correlation function.


It is unlikely that this incompleteness is responsible in itself for the dearth of low-redshift radio matches (cf. Section \ref{redshifts}), which make up a minority of the sources at $\sim$1 mJy. One would expect to preferentially lose the higher-redshift AGN at around the flux density limit. For example, the SKADS simulation features  $\sim$5 low-redshift star-forming galaxies per square degree compared to $\sim$30 AGN at higher redshifts, in the flux density range of 1--2 mJy.
	
\begin{figure*}
\centering
\includegraphics[width=0.94\textwidth]{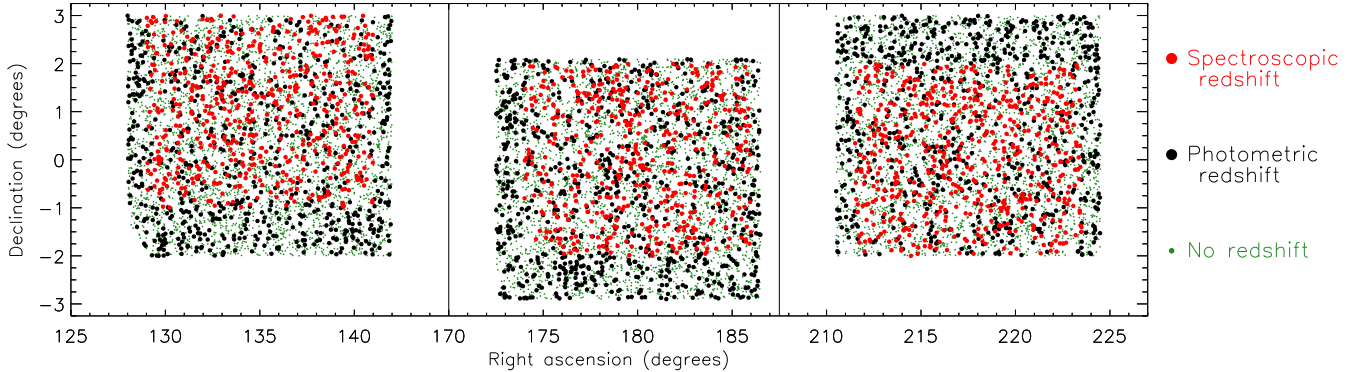}
\caption{The angular distribution of the radio samples used. Optically identified FIRST sources are plotted in red (GAMA spectroscopic redshifts) or black (SDSS/UKIDSS photometric redshifts), and those without optical matches are plotted in green.}
\label{samplearea}
\end{figure*}
	
\subsection{Sample Catalogues}

Table \ref{samples} lists the various samples to be investigated, comprising one large FIRST sample and 5 subsamples of FIRST sources within the GAMA fields. The large FIRST sample is over a single 115\degr $\times$ 64\degr patch (127.5\degr $<$ $\alpha$ $<$ 242.5\degr ; 0\degr $<$ $\delta$ $<$ 64\degr), while two smaller subsets contain all FIRST sources found within the enlarged GAMA footprint defined in Table 1 (FIRST(G)) and only those which are unmatched with GAMA sources (FIRST(G-unmatched)), shown in Figure \ref{samplearea}. The matched GAMA sources are also subsequently split into low- and high-redshift subsets in order to attempt to demonstrate any redshift evolution. The optical cross-matching preferentially identifies low-redshift radio sources, allowing us to use the unmatched sources to probe higher redshifts ($z_{\textrm{med}} = 1.55$) than the matched or full samples, further helping to observe any redshift-dependence of their clustering.

\begin{table}
\begin{center}
\caption{A summary of the samples for which the clustering properties will be measured: GAMA/SDSS cross-matched radio sources (further split into $z>0.5$ and $z<0.5$ samples), a large sample of FIRST sources, a smaller sample of all FIRST sources within the GAMA fields, and the remaining FIRST sources in these fields after removing GAMA cross-matches (all post-collapse of multiple sources). Redshift distributions are shown in Figure \ref{zdists}, and the enlarged GAMA area footprint is shown in Figure \ref{samplearea}.} 
\begin{tabular}{l c c c r} \hline
& Area & No. of & \multicolumn{2}{c}{Redshift} \\ \cline{4-5}
Sample & (deg$^2$) & sources &  Limit & $z_{\textrm{med}}$ \\ \hline
GAMA(matched) & 210 & 3,886 & - & 0.48 \\
" & 210 & 2,156 & $z<0.5$ & 0.30 \\ 
" & 210 & 1,730 & $z>0.5$ & 0.65 \\
FIRST & 5,922 & 342,615 & - & 1.21\\
FIRST(G) & 210 & 13,346 & - & 1.21 \\ 
FIRST(G--unmatched) & 210 & 9,460 & - & 1.55 \\ \hline
\end{tabular}
\label{samples}
\end{center}
\end{table}	

\section[]{Angular Correlation Function}\label{angular}

A simple measure of the clustering of sources in an angular distribution is found in the angular two-point correlation function, $w(\theta)$. It is defined as the excess probability of finding a galaxy at an angular distance $\theta$ from another galaxy, as compared with a Poissonian (unclustered) distribution \citep{peebles80}:
\begin{equation}
	\delta P = \sigma[1 + w(\theta)]\delta\Omega,
\end{equation}
where $\delta P$ is the probability, $\sigma$ is the mean surface density and $\delta\Omega$ is the surface area element.
	
A simple estimator defines $w(\theta) = DD/RR - 1$, where $DD(\theta)$ and $RR(\theta)$ represent the number of galaxy pairs separated by $\theta$ in the real data and a random catalogue, respectively. Introducing an additional count of the cross-pair separations $DR(\theta)$ allows for reduced variance, such as is found for our chosen estimator by \citet{landy93}:
\begin{equation}
w(\theta) = \frac{n_r (n_r - 1)}{n_d (n_d - 1)} \frac{DD}{RR} - \frac{(n_r - 1)}{2n_d} \frac{DR}{RR} + 1,
\end{equation}	
where $n_d$ and $n_r$ are the number of real sources and random sources.	

By averaging over several random data sets and using $\overline{DR}$ and $\overline{RR}$ or by using a more densely populated random catalogue, we may assume the statistical error in the random sets to be negligible. The random catalogues themselves have been autocorrelated finding no significant deviation from zero even at extremes of angular separation where bin counts are lowest. The uncertainty on $w$, therefore, is often given by the Poisson error due to the DD counts alone
\begin{equation}
	\Delta w = \frac{1 + w(\theta)}{\sqrt{DD}}.
\end{equation}

However, the errors in the correlation function depend on the $DD$ counts beyond simple Poisson variance; adjacent bins are correlated, with each object contributing to counts across a range of separation bins. The errors are therefore calculated somewhat more rigorously using a bootstrap resampling technique \citep{ling86} whereby several data catalogues are constructed by randomly sampling (with replacement) the original set of objects. As such, in any given set, some sources are counted twice or more and some not at all. The resulting binned $DD$ counts should give a mean approximately equal to the original data but allow us to calculate a variance for each bin, and therefore $w(\theta)$ values. \citet{cress96}, for example, found errors in $w(\theta)$ for the early FIRST survey with a bootstrap resampling method. They found the Poisson error estimates to be too small by a factor of 2 on small scales ($\sim$ 3\arcmin) and more than an order of magnitude for larger scales ($\sim$ 5\degr). We find Poisson error estimates to be consistently a factor of two smaller than the bootstrap error up to $\sim$ 1\degr, above which the ratio increases approximately exponentially.

The restricted survey area from which we can measure $w(\theta)$ also results in a negative offset in the observed correlation function, known as the \emph{integral constraint}. Expressed mathematically, the relation between the observed correlation function $w_\textrm{obs} (\theta)$ and the genuine function $w(\theta)$ is
\begin{equation}
w_\textrm{obs} (\theta) = w(\theta) - \sigma^2,
\end{equation}
where $\sigma^2$ represents the integral constraint \citep{groth77} which can be approximated, following \citet{roche99}, by
\begin{equation}
\sigma^2 = \frac{\sum RR(\theta) w(\theta)}{\sum RR(\theta)} .
\end{equation}

\begin{figure*}
\centering
\begin{subfigure}{0.4\textwidth}
\includegraphics[width=\textwidth]{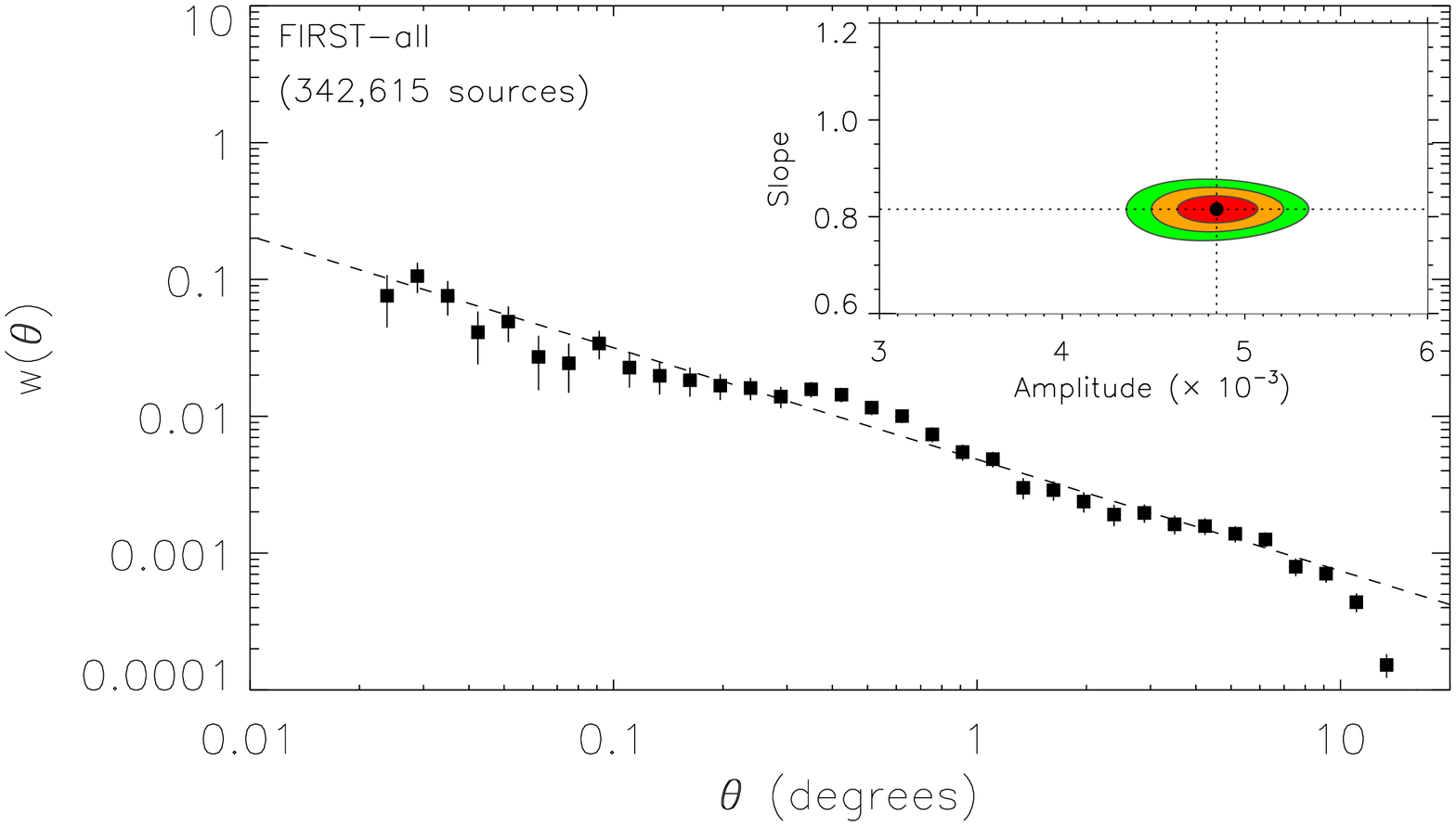}
\end{subfigure} \qquad
\begin{subfigure}{0.4\textwidth}
\includegraphics[width=\textwidth]{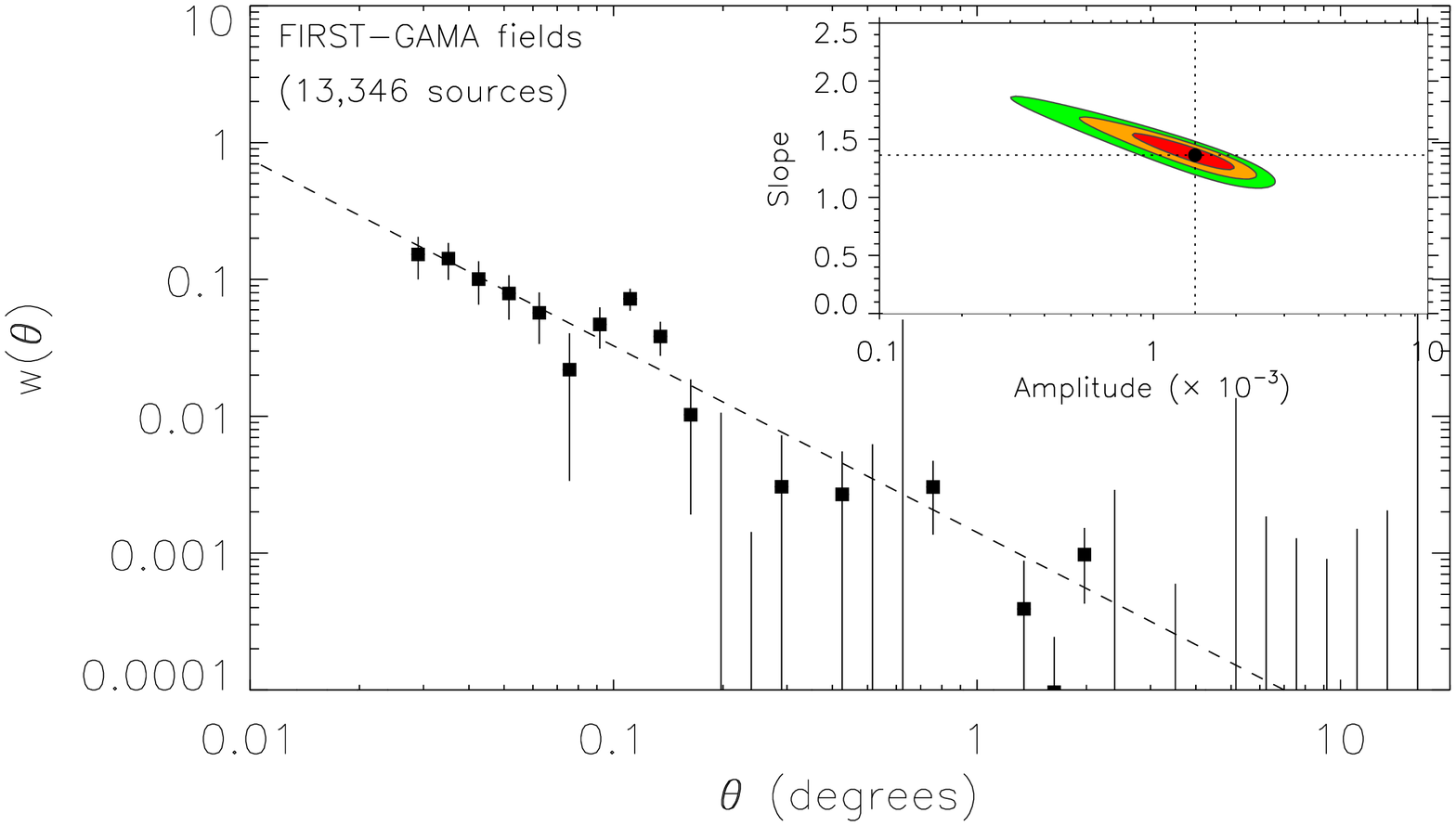}
\end{subfigure} \\
\begin{subfigure}{0.4\textwidth}
\includegraphics[width=\textwidth]{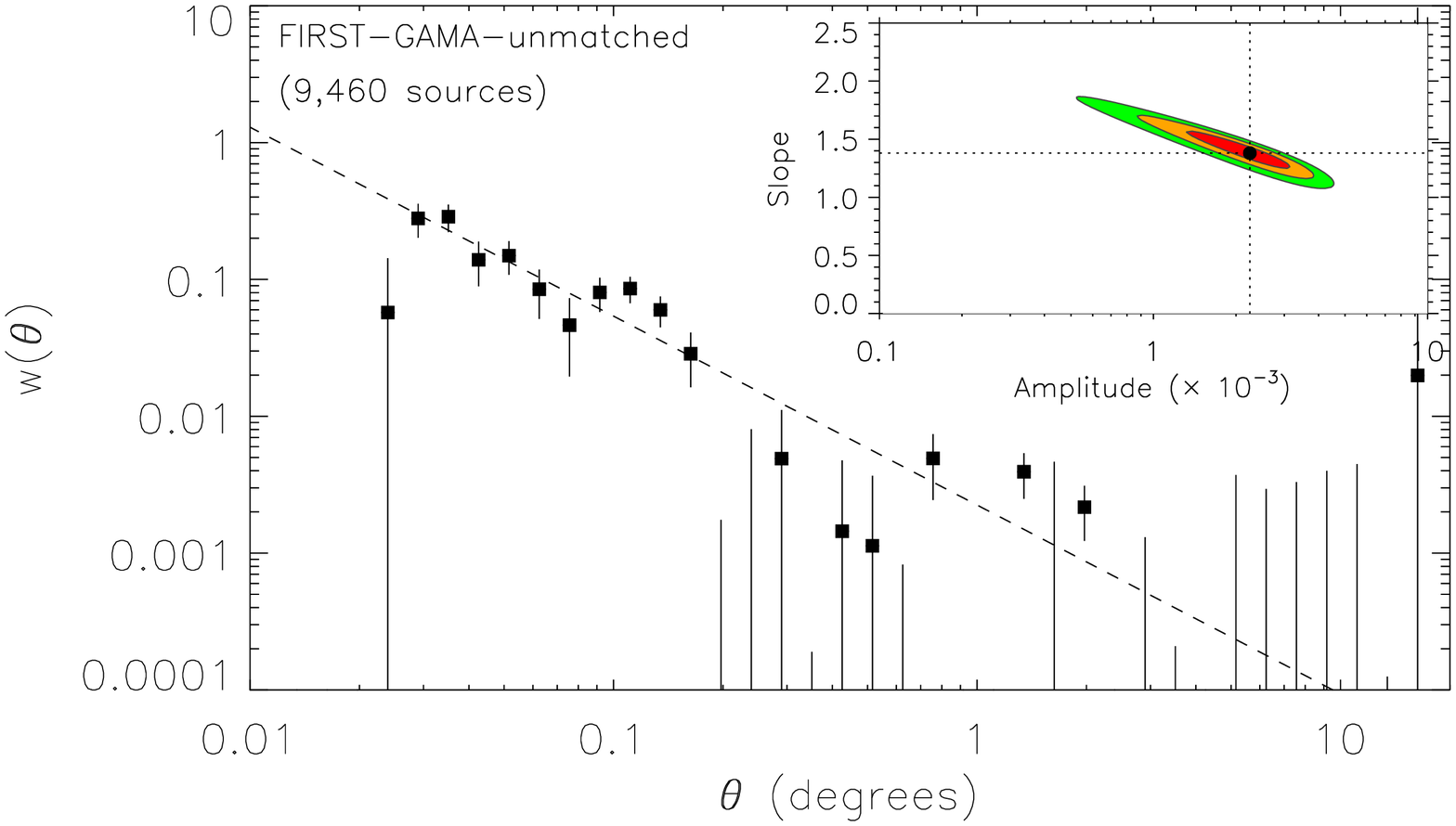}
\end{subfigure} \qquad
\begin{subfigure}{0.4\textwidth}
\includegraphics[width=\textwidth]{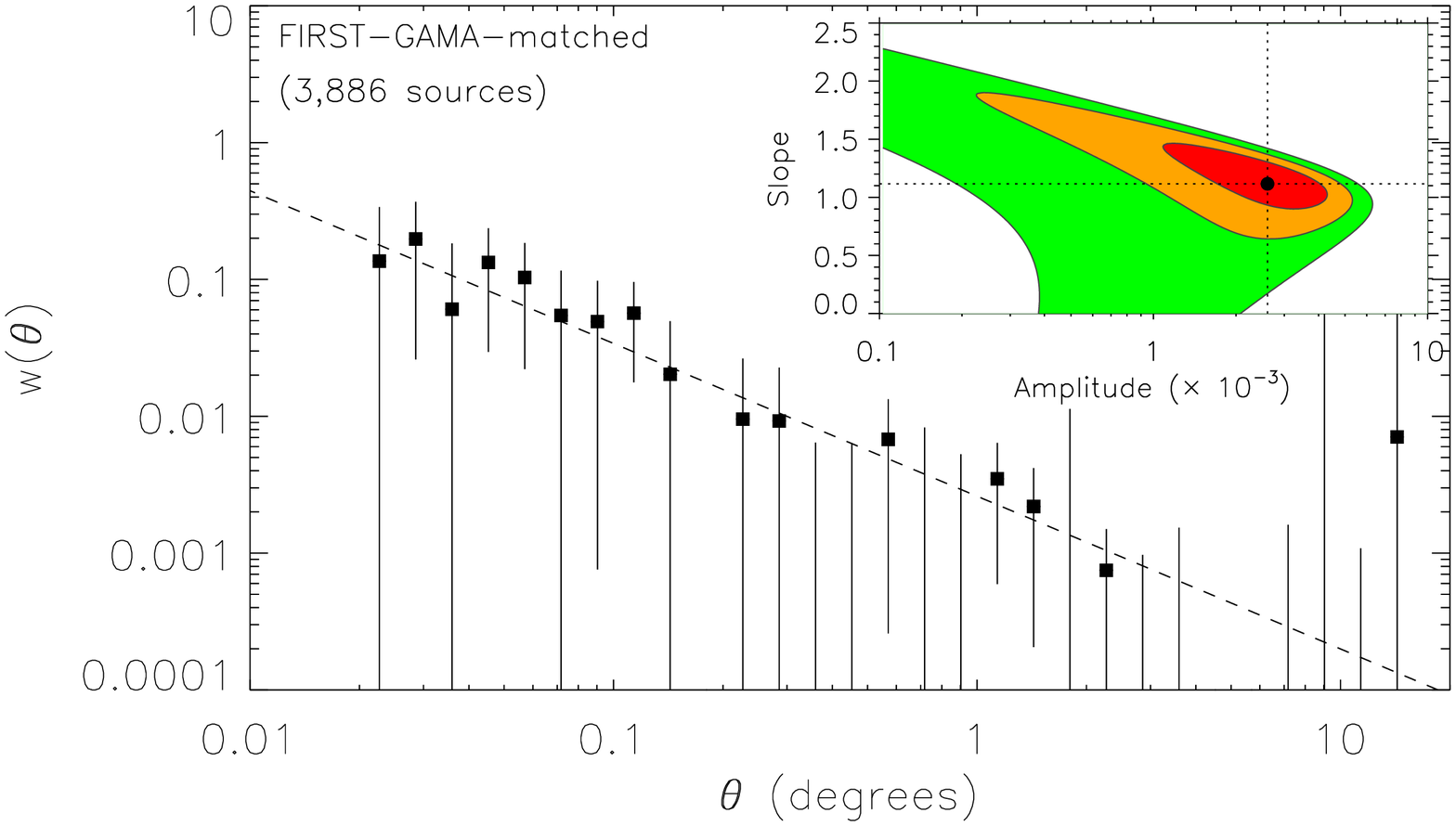}
\end{subfigure} \\
\begin{subfigure}{0.4\textwidth}
\includegraphics[width=\textwidth]{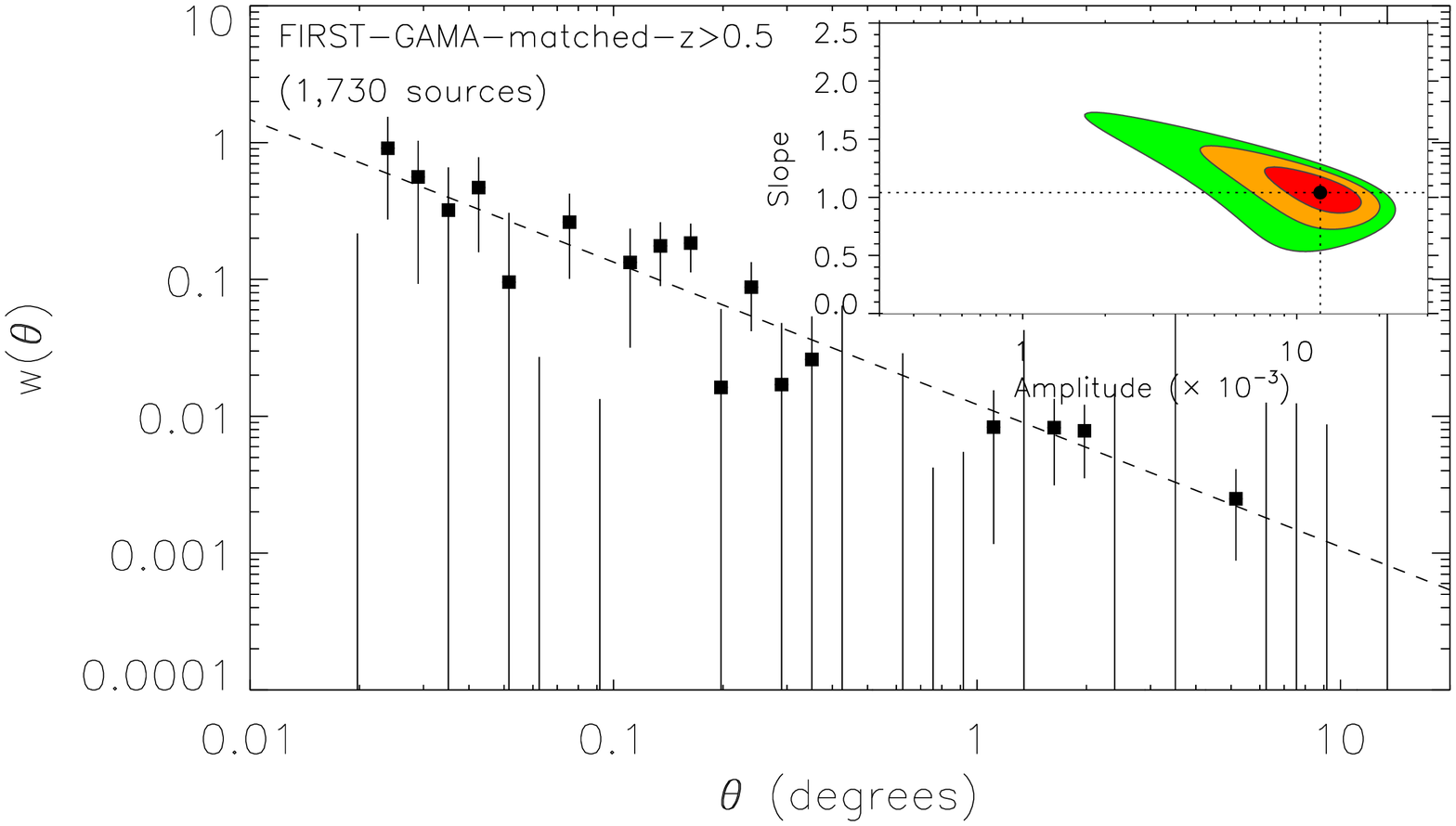}
\end{subfigure} \qquad
\begin{subfigure}{0.4\textwidth}
\includegraphics[width=\textwidth]{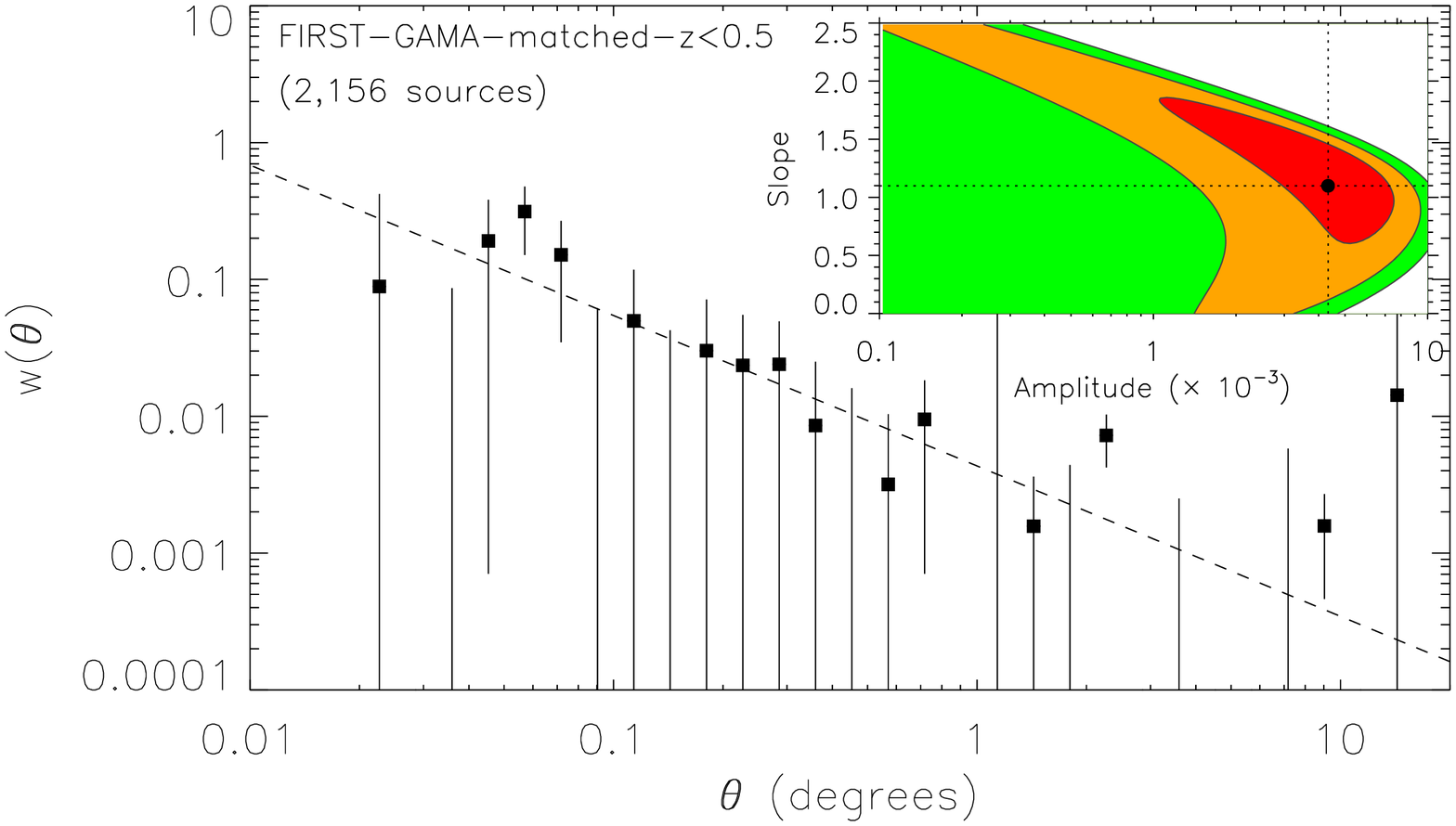}
\end{subfigure} 
\caption{The angular correlation function for the 5 samples (sample name is denoted at the top left corner of each panel) within the extended GAMA fields (with bootstrap resampling errors) and the larger FIRST sample in the bottom right panel. The dashed lines show the best fit power-law (over the range $0.02 < \theta < 10$ degrees) and the inset contour plots shows $\chi^2$ parameter fits at 68, 90 and 95 per cent confidence levels.}
\label{GAMAw}
\end{figure*}

Traditionally, $w(\theta)$ has been fitted by a power law (e.g. Peebles, 1980) with a slope of $\sim 0.8$ commonly found for the clustering of objects of various masses \citep{bahcall83}. While radio sources have been found to fit a distinct double power law (cf. \citealt{blake02a,overzier03}), since we have collapsed the multiple component sources, $w(\theta)$ should reduce to the canonical single power law form for our data.
			
We fit $w(\theta)$ with a single power law function of the form $w(\theta) = A \theta^{1-\gamma}$, fitted over the range $0.02 < \theta < 10$ degrees. This is done using the Metropolis-Hastings algorithm to obtain a Markov Chain Monte Carlo (MCMC) simulation of $10^6$ data points in $\gamma$--log(A) space. Parameter values quoted are minimum-$\chi^2$ values and 1$\sigma$ errors correspond to the region containing 68.3 per cent of the MCMC points. A Levenberg-Marquardt $\chi^2$ minimization routine yields the same best fit values, and ordinary $\chi^2$ contours coincide closely with those from the MCMC simulations. 	
	\vspace{5 mm}
	
Figure \ref{GAMAw} shows the results of this angular correlation function method for the 5 samples of FIRST radio sources within the GAMA regions, and we see clearly in the likelihood contours that the errors are smaller for the more numerous unmatched sources than the matched sources. The best fit parameters are shown in Table \ref{parameters}). The unmatched sample has a lower amplitude than the matched sample, but given these will come to represent complementary high- and low-redshift measurements, respectively, similar amplitudes do not imply similar clustering scales due to the increased angular diameter distance at high redshift.

\section[]{Limber Inversion}\label{Limber}

	The 3-dimensional analogue of $w(\theta)$ is the spatial two-point correlation function $\xi(r)$, which measures the excess probability, due to clustering, of finding a pair of objects separated by $r\rightarrow r + \delta r$ as compared with a Poissonian (unclustered) distribution, defined as:
\begin{equation}
	\delta P = n[1 + \xi(r )]\delta V,
\end{equation}
where $n$ is the mean number den	sity of objects and $\delta V$ the volume element.

If the redshift distribution of a set of objects is known, one may deproject the angular correlation function into the spatial correlation function. This is the purpose of the cosmological Limber equation \citep{limber53,peebles80} for estimating the spatial correlation length, $r_0$ (discussed further in Section~\ref{spatial}). This is often more useful than computing the spatial correlation function, $\xi(r )$, directly as a complete set of individual redshifts is rarely available for a given survey, thus requiring the redshift distribution to be estimated via the luminosity function in order to deproject $w(\theta)$. Using the redshifts available for the objects in our catalogue, however, means we may apply a distribution directly from the data for the radio sources with optical counterparts. 

An epoch-dependent form of the spatial correlation function is assumed \citep[see e.g.][and references therein]{dezotti90,overzier03}:
\begin{equation}
	\xi(r,z) = \left(\frac{r_0}{r}\right)^\gamma \times (1+z)^{\gamma-(3+\epsilon)} ,
\end{equation}
where $r$ is in comoving units and $\epsilon$ parameterises the clustering model being assumed. \citet{overzier03} and \citet{kim11} offer 3 main models: \emph{stable clustering} (where clusters have fixed physical size; $\epsilon=0$), \emph{comoving clustering} (where clusters have fixed comoving size; $\epsilon=\gamma - 3$) or \emph{linear clustering} (growth under linear perturbation theory; $\epsilon = \gamma - 1$). For a typical slope found in the literature of $\gamma \sim 2$, these clustering models are stable, decaying and growing, respectively. Other authors, such as \citet{elyiv12}, also apply $\epsilon=-3$. Given that $\epsilon > 0$ (or $<0$) implies growing (or decaying) clustering, this final $\epsilon$ value implies a more rapid clustering decay than the other models. In this paper we adopt $\epsilon=\gamma-3$ and $\epsilon=\gamma-1$ to provide a conservative range of values at high redshift

The spatial correlation function slope, $\gamma$, is the same as that used in the power law fit to the angular correlation function (where the magnitude of the slope is $\gamma - 1$), so we measure this parameter through the $w(\theta)$ function. The amplitude $A$ of $w(\theta)$ has been expressed as a function of $r_0$ (in comoving coordinates) in the literature \citep{overzier03,kovac07,kim11,elyiv12} as follows:
\begin{equation}
A = r_0^\gamma H_\gamma \left(\frac{H_0}{c}\right) \frac{\int_0^\infty N^2(z) (1+z)^{\gamma - (3+\epsilon)} \chi^{1-\gamma}(z) E(z) \mathrm{d}z}{\left[ \int_0^\infty N(z) \, \mathrm{d}z \right]^2},
\label{eq:Limber}
\end{equation}
where  $H_\gamma$ is related to the Gamma function, $H_\gamma =  \Gamma(\frac{1}{2}) \Gamma(\frac{\gamma-1}{2})/\Gamma(\frac{\gamma}{2})$, $N(z)$ is the redshift distribution and $\chi(z)$ is the comoving line-of-sight distance to an object at a redshift $z$:
\begin{equation}
\chi (z) = \frac{c}{H_0} \int^{z}_{0} \frac{\mathrm{d}z^{\prime}}{E(z^{\prime})}.\label{eq:comoving}
\end{equation}
Here, $H_0$ is the Hubble constant and $E(z)$ is the function used to describe the cosmological expansion history:
\begin{equation}
E(z) = \left[\Omega_{m,0} (1+z)^3 + \Omega_{k,0} (1+z)^2 + \Omega_{\Lambda,0}\right]^\frac{1}{2}. \label{eq:E(z)}
\end{equation} 
Equation \ref{eq:Limber} may simply be inverted to give the comoving correlation length, $r_0$ as a function of (i) redshift distribution, (ii) correlation function slope and (iii) angular clustering amplitude.

We employ redshift distributions as measured for the GAMA matched sources, and for the simulated SKADS sources, as detailed in Section \ref{redshifts}. We also emphasise that using the SKADS $N(z)$ allows us to account for sensitivity variations across the FIRST survey as we able to remove the correct number of sources of a given flux and redshift (as described in Section \ref{completeness}).

\subsection{Results of Limber Inversion}

\renewcommand{\arraystretch}{1.3}
\begin{table*}
\begin{center}
 \caption{Clustering parameters measured covering real and simulated radio sources ($S_{1.4}>1$ mJy) with real (GAMA) and/or simulated redshift distributions (FIRST). Errors are $1\sigma$ from $10^6$ ($A$, $\gamma$) and $10^5$ ($r_0$, $b$) MCMC calculations.} 
  \begin{tabular}{l c c c c c c c c c} \hline
  & & & & & \multicolumn{2}{c}{$r_0$ ($h^{-1}$Mpc)} & & \multicolumn{2}{c}{$b(z=z_{\textrm{med}})$} \\ \cline{6-7} \cline{9-10}
   Sample & $N_{obj}$ & A ($\times 10^{-3}$) & $\gamma$ & $z_{\textrm{med}}$ & $\epsilon = \gamma-1$ & $\epsilon = \gamma-3$ && $\epsilon = \gamma-1$ & $\epsilon = \gamma-3$ \\ \hline
   FIRST--all & 342,615 & $4.87^{+0.12}_{-0.17}$ & $1.82^{+0.02}_{-0.02}$ & 1.21 & $11.82^{+0.43}_{-0.46}$ & $8.20^{+0.41}_{-0.42}$ && $4.39^{+0.19}_{-0.19}$ & $3.14^{+0.16}_{-0.17}$\\	
   FIRST--GAMA fields & 13,346 & $1.31^{+0.35}_{-0.34}$  & $2.35^{+0.11}_{-0.09}$ & 1.21 &$11.27^{+0.82}_{-0.87}$ & $10.53^{+0.64}_{-0.68}$ && $6.40^{+0.42}_{-0.45}$ & $5.95^{+0.48}_{-0.53}$\\  
   FIRST--GAMA--unmatched & 9,460 & $2.21^{+0.62}_{-0.57}$ & $2.39^{+0.11}_{-0.10}$ & 1.55 & $14.41^{+1.23}_{-1.46}$ & $13.60^{+0.83}_{-1.07}$ &&  $10.06^{+0.49}_{-0.50}$ & $9.45^{+0.58}_{-0.67}$\\
   FIRST--GAMA--matched & 3,886 & $2.76^{+0.94}_{-1.11}$ & $2.15^{+0.12}_{-0.24}$ & 0.48 & $8.24^{+1.75}_{-2.36}$ & $6.72^{+1.81}_{-2.17}$ && $2.65^{+0.98}_{-0.90}$ & $2.13^{+0.90}_{-0.76}$ \\
   FIRST--GAMA--matched--z$<$0.5 & 2,156 & $4.28^{+1.99}_{-1.89}$ & $2.28^{+0.24}_{-0.51}$ & 0.30 & $6.21^{+2.34}_{-4.34}$ & $5.39^{+2.53}_{-3.92}$ && $1.77^{+1.62}_{-1.23}$ & $1.52^{+1.57}_{-1.07}$ \\ 
   FIRST--GAMA--matched--z$>$0.5 & 1,730 & $12.26^{+2.90}_{-3.04}$ & $2.04^{+0.12}_{-0.12}$ & 0.65 & $17.07^{+4.50}_{-4.70}$ & $10.67^{+3.22}_{-3.18}$ && $5.74^{+2.56}_{-1.97}$ & $3.56^{+1.59}_{-1.23}$\\  \hline
  \end{tabular}
   \label{parameters}
\end{center}
\end{table*}

\begin{figure*}
\begin{minipage}[b]{0.48\linewidth}
\centering
\includegraphics[width=\textwidth]{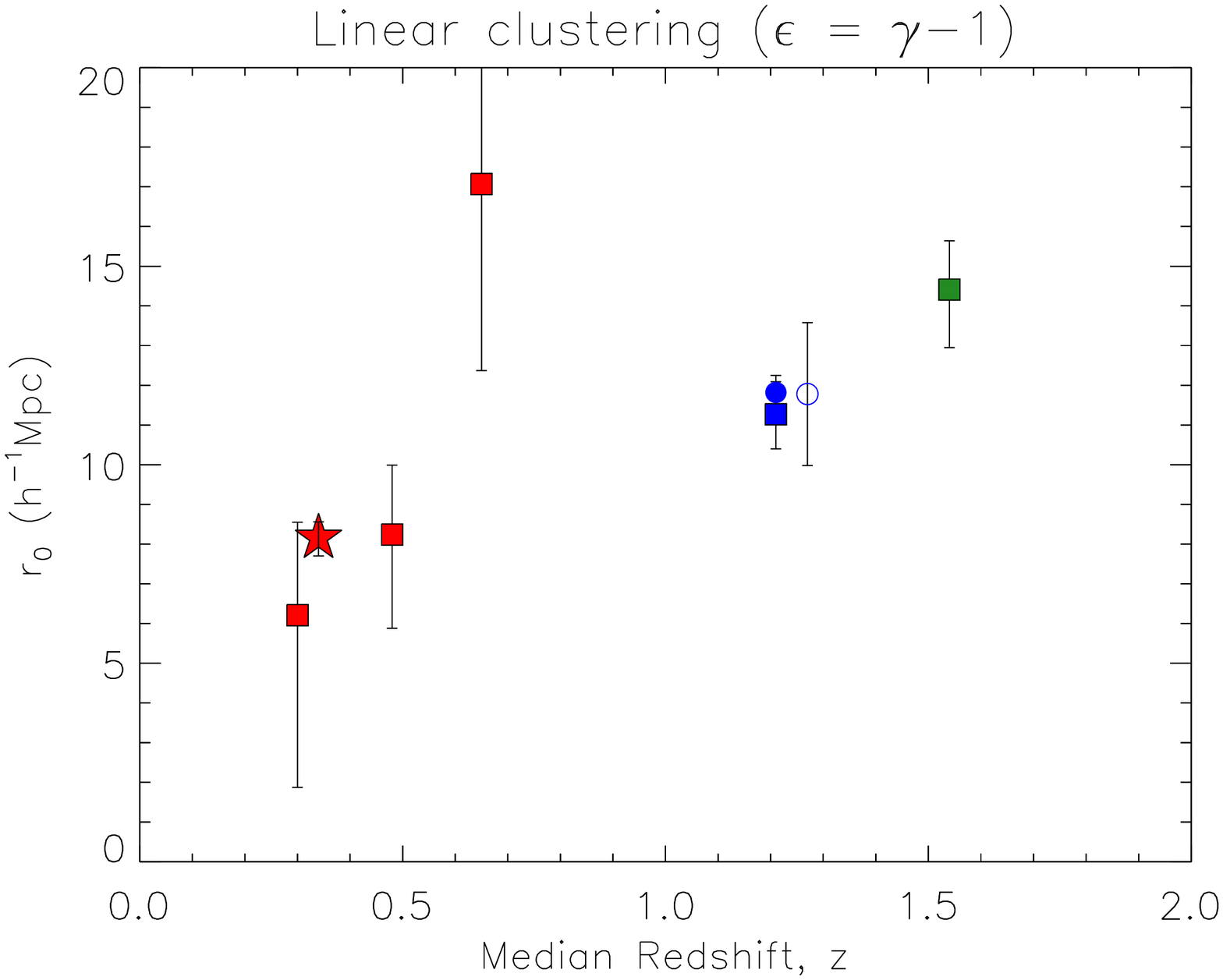}
\end{minipage}
\begin{minipage}[b]{0.48\linewidth}
\centering
\includegraphics[width=\textwidth]{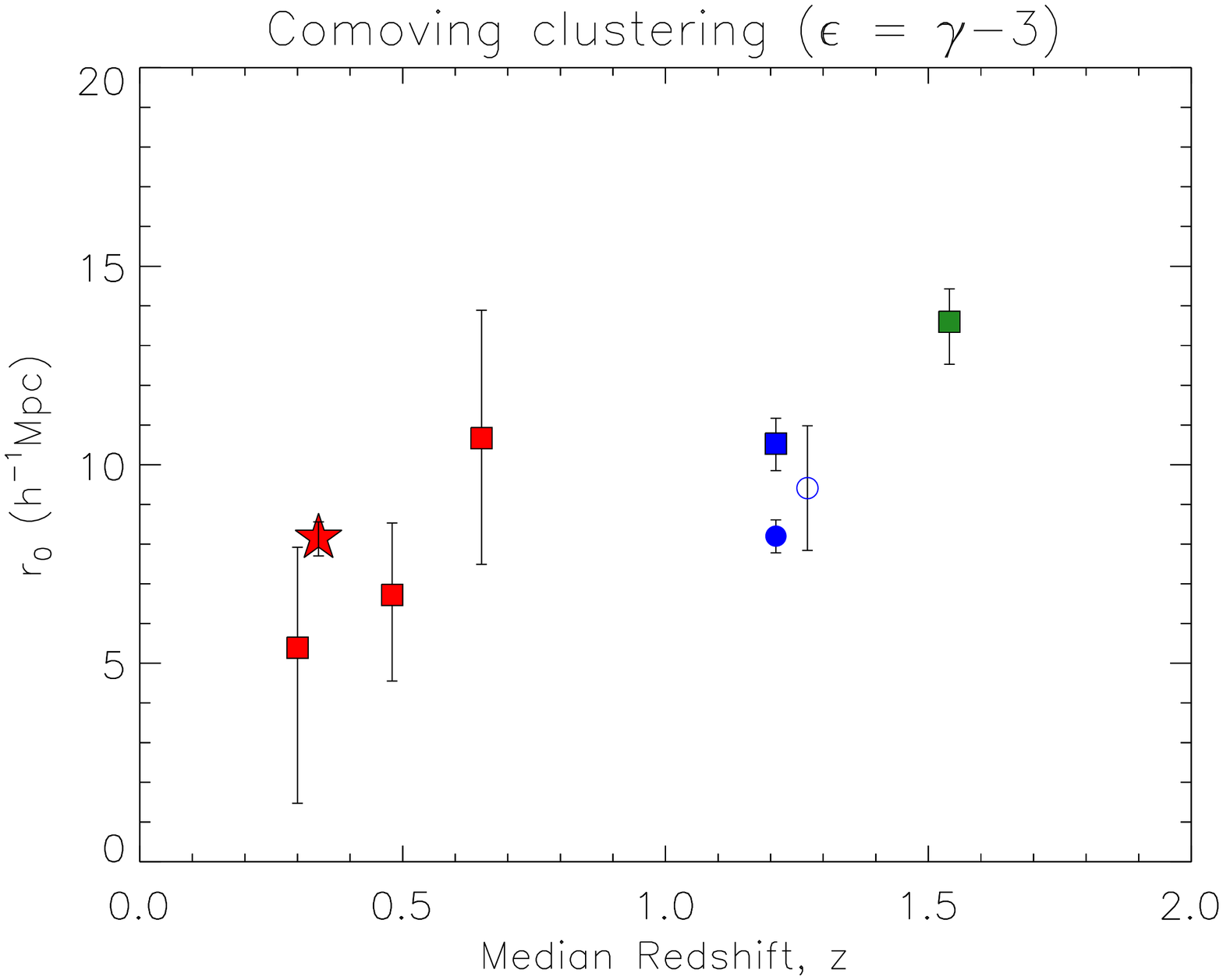}
\end{minipage}
  \caption{Correlation length for the observed radio samples described in Table \ref{parameters} as calculated for two different clustering indices. Filled square symbols correspond to samples over the GAMA/SDSS/UKIDSS fields, while the circle symbols refer to the wider FIRST survey, and the star symbols correspond to the spectroscopic sample of FIRST matches in GAMA. Red, green and blue colours refer to matched, unmatched and total samples respectively (cf. Figure \ref{zdists}). The open circle beside the points at $z\sim1.2$ shows the mean value and error associated with 20 independent patches of the FIRST survey, providing an estimate of the sample variance within a field size of 210 square degrees.}
  	\label{r0}
\end{figure*}

We have measured $w(\theta)$ for the six samples described above, and fitted them with a power law (see Table \ref{parameters}). Given the dependence on $\epsilon$, the clustering index of choice, $r_0$ is shown for the comoving ($\epsilon = \gamma - 3$) and linear ($\epsilon = \gamma -1$) clustering models. The latter consistently gives lower values for $r_0$, with the unshown stable ($\epsilon = 0$) model values being between the two, but these differences tend to be comparable with the associated errors. The results of the comoving clustering model will be quoted hereafter, unless stated otherwise.

The large FIRST sample yields a $w(\theta)$ consistent with the literature, where we find a slope of $\gamma - 1 = 0.82\pm0.02$, the collapsing of close pairs successfully having removed evidence of a steeper, small-scale power law component. By assuming the same redshift distribution as a similar SKADS catalogue (adjusted in accordance with the noise variations in the FIRST survey, see Sec.~\ref{completeness}), we find a correlation length of $8.20^{+0.41}_{-0.42}$ $h^{-1}$Mpc at a median redshift of $z = 1.21$ 

We are able to infer the clustering properties from our real data within the GAMA fields at 3 different epochs utilising the three subsamples described in Section~\ref{redshifts} and Figure \ref{zdists}. 
These 3 subsamples have distributions of median $z=$ 0.48, 1.21 and 1.55, for the matched, complete, and unmatched radio samples, respectively. 

In each case, the $w(\theta)$ power law is found to be significantly steeper ($\gamma>2$) than for the larger FIRST sample, albeit less well constrained owing to the smaller sample sizes. This steeper power law is borne out in the analysis where we find that the ``FIRST-GAMA fields'' sample has a larger clustering length than the larger FIRST sample with the same $N(z)$. The trend in these 3 subsamples, however, is for increasing $r_0$ with redshift. This is confirmed by analysing subsamples of the cross-matched objects. Small number counts limit us to taking a simple high- and low-redshift approach, but even within this redshift range of $z\sim0.3$ to $z\sim0.65$, $r_0$ is seen to increase.

In order to assess the effect of cosmic variance on our results over the GAMA fields, we also find $w(\theta)$ and $r_0$ for 20 separate 210 square degree patches of the FIRST survey over a large range of declinations ($-5\degr < \delta < +55\degr$), assuming the SKADS $N(z)$ for each. We find a mean and standard deviation of correlation lengths from these subsamples to give $r_0 = 9.41 \pm 1.57$ $h^{-1}$Mpc, lying between the two values found for our GAMA sample and for the larger FIRST sample at $z = 1.21$, and placing both within this cosmic variance error.

\subsection{Discussion}
Previous results using the angular clustering approach, without individual redshifts, \citet{blake02a} find a clustering length for $S>10$ mJy NVSS sources of $r_0 \sim 6$ $h^{-1}$Mpc (independent of flux density limit) under the assumptions of $N(z)$ calculated from luminosity functions of \citet{dunlop90} and an Einstein-de Sitter (EdS) cosmological model. Similarly, \citet{overzier03} use NVSS and similarly derived $N(z)$ from \citet{dunlop90} finding $r_0 \sim 5 \pm 1$ $h^{-1}$Mpc  for flux density between 3 and 40 mJy. Both papers probe clustering at $z \sim 1$ and, while they differ slightly from one another, they are both considerably lower than our findings even with our lower flux limit diluting the clustering due to AGN. It is important to note, however, that the Dunlop \& Peacock models used are poorly constrained at these low flux density limits due to a lack of volume coverage at low redshift in their combined surveys. Improvements in deep surveys have allowed \citet{wilman08} to better constrain low-power AGN and star-forming galaxies in the SKADS redshift distribution for use at the mJy level.

Any underestimation of $r_0$ could partly be a result of having fixed the angular clustering power law to a slope of $-0.8$ as is widely observed for normal galaxies, although even our larger FIRST sample with $\gamma - 1 = 0.77$ appears to exhibit stronger clustering. This may be simply to reduce the power law fitting to a one-parameter problem, but a steeper slope ($\sim$ 1.0 to 1.4) may give a higher correlation length than found under the assumption of a universal slope of $-0.8$. It is likely, however, that these differing results are due mostly to our lower flux density limits and slightly higher median redshift.

Indeed, while our $w(\theta)$ fit for the larger FIRST sample does yield the assumed $(\gamma -1) = 0.8$, \citet{cress96} find a higher value, similar to our smaller subsamples, of $(\gamma - 1) \sim 1.1$ yielding a larger correlation length at the same $z \sim 1$ of $r_0 \sim 10$ $h^{-1}$Mpc, more in keeping with our results. Furthermore, this epoch and corresponding correlation length is in close agreement with observations of $z \sim 1$ extremely red objects (EROs) using the same $\epsilon = \gamma - 3$ clustering model (e.g. \citealt{daddi01,daddi02, mccarthy01,roche02}). This ties in with the suggestion by \citet{willott01} that EROs and high-redshift galaxies are different evolutionary stages of the same galaxy population, which in turn may be the progenitors of local bright ellipticals. 

Two studies of spectroscopic samples of radio sources associated with luminous red galaxies (LRGs) by \citet{wake08} and \citet{fine11} investigate samples at a similar range of redshifts below $z=0.8$ by different methods. \citet{wake08} fit a halo occupation distribution (HOD) model to the projected spatial correlation function of radio-detected ($L_{1.4} \gtrsim 10^{24.2}$ WHz$^{-1}$) LRGs at $z \sim 0.55$ to give $r_0 \sim 12$ $h^{-1}$Mpc, slightly greater than we find (Figure \ref{r0}) due to their selection of the most optically luminous, more powerful AGN sources. \citet{fine11} show the evolution in this clustering by adding LRG samples from SDSS ($z\sim0.35$) and AAOmega ($z\sim0.68$). Over this very similar redshift range to our optically-identified radio sources, these authors find $r_0$ using angular auto- and cross-correlation functions which are consistent with no redshift evolution, falling in the range of $\sim$10--12 $h^{-1}$Mpc. While we find a similar correlation length towards higher redshift, this result is contrary to the trend we observe with the GAMA/SDSS radio sources. Again, this could simply be a result of the different optical selection favouring considerably more powerful AGN at lower redshift, whereas our magnitude limit allows more of the low optical luminosity sources into our sample.

Looking at the AGN population from the other end of the electromagnetic spectrum, \citet{elyiv12} use a $N(z)$ derived from a range of luminosity functions for AGN detected in X-rays with the XMM-LSS, recovering the ubiquitous slope of $-0.8$ for the soft band sources, but $-1.0$ for the hard band. Their hard band sources exhibit a clustering length of $r_0 \simeq 10 \pm 1$ $h^{-1}$Mpc at $z \sim 1$ using a clustering index of $\epsilon = -1.2 \simeq \gamma - 3$ roughly corresponding to comoving clustering evolution. Accounting for the slightly lower redshift, this agrees well with a FIRST radio population expectedly dominated by AGN emission.


\begin{figure}
\centering
\includegraphics[width=0.48\textwidth]{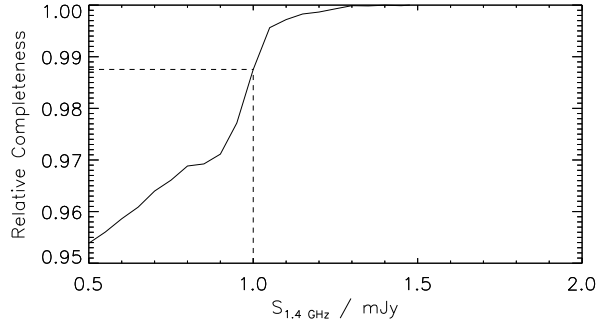}
\caption{Estimated ratio of completeness curves for the FIRST survey over the 3 equatorial GAMA/SDSS fields and a large area of the Northern sky. Higher rms noise values at lower declinations make the GAMA fields less complete at the lowest flux densities.}
\label{fig:completeness}
\end{figure}

\subsubsection{Surface Density Variations and Sample Variance}\label{variance}


\citet{blake02b} have shown some declination dependence in the surface density of FIRST sources, with a dearth at lower declination of fainter radio sources in particular. We find this to be an 8--10 per cent underdensity for sources fainter than 10 mJy with $|\delta|<3^{\circ}$, a range encompassing the GAMA fields. Given that the source densities remain consistent with expectations from SKADS, this density variation may be a result of increased noise in this region of the survey, sample variance, or interferometric effects arising from observing at low declination. Accounting for any variation in the noise values across the FIRST survey, however, finds only a small difference ($\sim$1 per cent at the flux density limit) in completeness of the equatorial fields and the wider survey (see Figure \ref{fig:completeness}). 

 As a result, we might expect the clustering statistics to reflect a larger proportion of radio-loud AGN rather than low-redshift star-forming galaxies which become more dominant at the mJy level. We demonstrate this effect by calculating $r_0$ and bias for 20 patches of FIRST, each 210 square degrees in area, over a declination range between $-6$\degr\,and $+54$\degr. The open circles in Figure \ref{r0}, slightly offset in $z$ for clarity, represent the mean and standard error on these quantities. The GAMA fields yield results at the high end of what bias we might find based on the surface density fluctuation, while the larger FIRST sample at $\delta > 0$ is towards the lower end, however we note that both are consistent with the expected variance with patches of 210 square degrees.

\section[]{Spatial Correlation Function}\label{spatial}

In addition to calculating the spatial correlation length to high redshift using the deprojected angular correlation function, we are also able to determine the spatial correlation function directly for those sources with spectroscopic redshifts.
Unlike the angular case, calculating spatial separations between galaxies requires information about the underlying cosmology and is dependent on the dynamics of the clustered systems. On face value, the only extra data required for our catalogue is a line-of-sight distance, which we calculate based on observed redshifts. Given the comoving distances to two objects, $\chi_1$ and $\chi_2$ (from Equation \ref{eq:comoving}), and their angular separation, $\theta$, their comoving spatial separation is given by
\begin{equation}
r = \left({\chi_1}^2 + {\chi_2}^2 - 2\chi_1 \chi_2 \cos{\theta} \right)^{\frac{1}{2}} ,
\label{eq:separation}
\end{equation}
assuming a flat cosmology (see \citealt{liske00} for more general expressions).

The spatial two-point correlation function is usually fitted by a single power law over a significant range of separations:
\begin{equation}
	\xi(r,z)  = \left(\frac{r_0(z)}{r}\right)^{\gamma} \label{eq:xir} ,
\end{equation} 
where $r_0 (z) = r_0 (1+z)^{1-\frac{3+\epsilon}{\gamma}}$, incorporating the clustering evolution described in Section \ref{Limber}. 

\begin{figure*}
	\centering
	\includegraphics[width=0.48\textwidth]{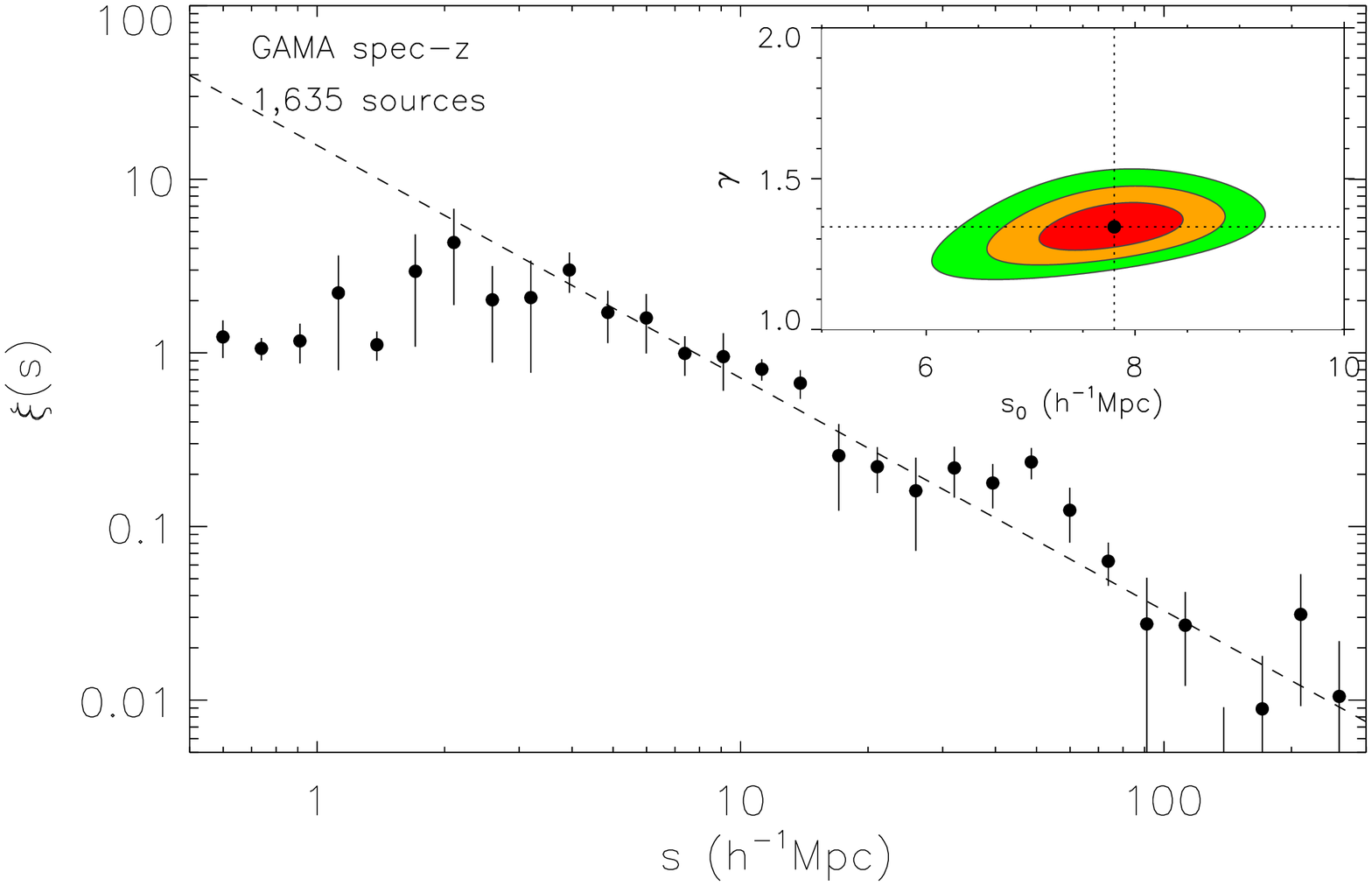}
	\hspace{5pt}
	\includegraphics[width=0.48\textwidth]{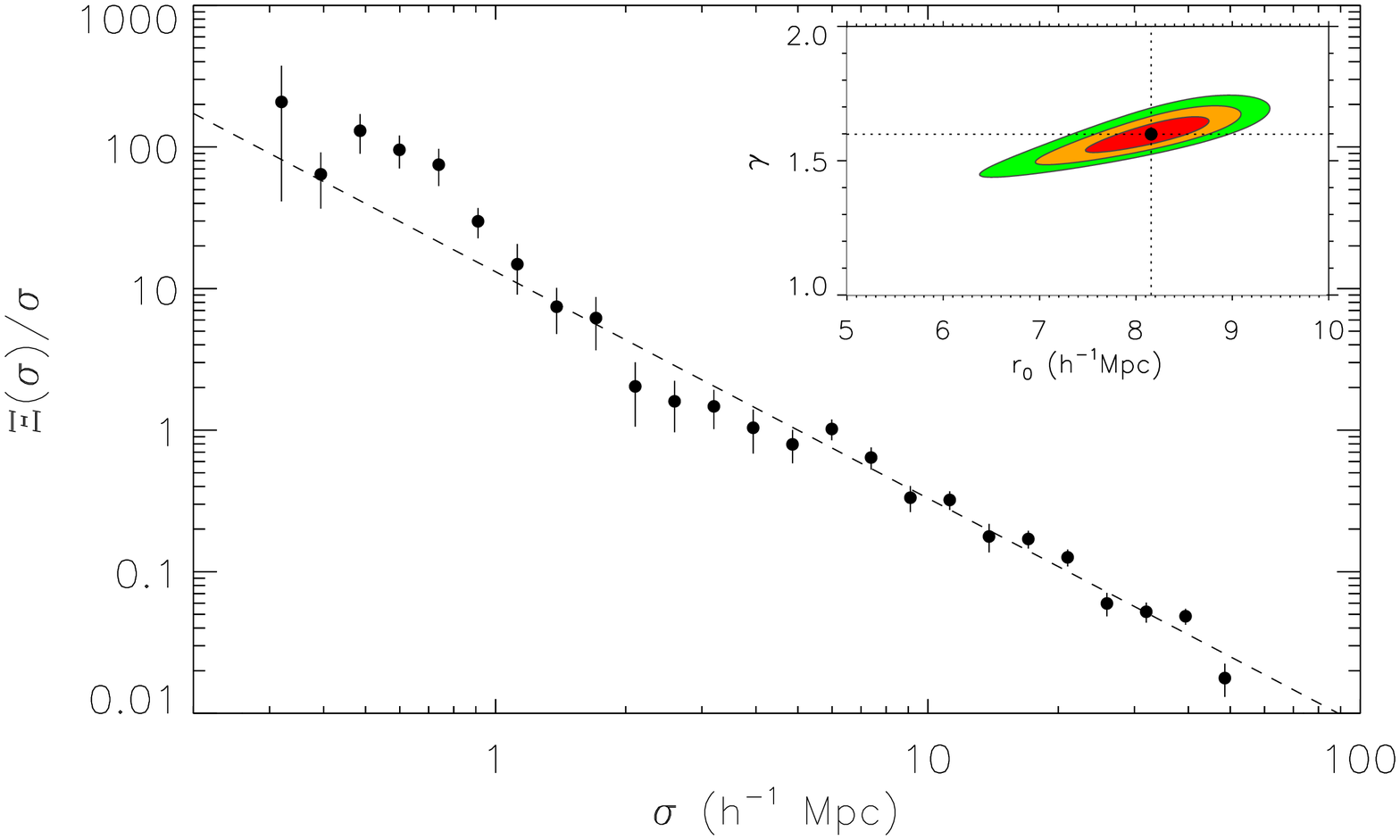}
	\caption{The redshift-space correlation function (\textit{left}) and projected spatial correlation function (\textit{right}) for the GAMA spectroscopically identified objects with bootstrap resampling errors. The dashed line shows the best fit power-law (over the ranges $2 < s < 300$ $h^{-1}$Mpc and $\sigma < 50$ $h^{-1}$Mpc, respectively) and the inset plot shows $\chi^2$ parameter fits at 68, 90 and 95 per cent confidence levels.}
	\label{fig:xis}
\end{figure*}

%
%
\subsection{Projected Correlation Function}\label{projxi}

By calculating galaxy distances from their redshifts and a cosmological model as in Equations \ref{eq:comoving} and \ref{eq:separation}, we actually find the {\it redshift-space} correlation function, $\xi(s)$, which is systematically different from the true {\it real-space} correlation function $\xi(r)$. The difference occurs because measured redshifts are not entirely the result of cosmological expansion but they are also affected by radial peculiar velocity components of the source distorting the observer's view (the Kaiser effect; \cite{kaiser87} and ``Fingers of God''). These effects make it difficult to find the real-space clustering parameters $r_0$ and $\gamma$ from direct $\xi(s)$ measurements. These can be found directly through the projected correlation function $\Xi(\sigma)$ or indirectly through deprojecting the angular correlation function $w(\theta)$ (Section \ref{Limber}).

\begin{table}
\begin{center}
\caption{Clustering parameters found from the redshift-space correlation function $\xi(s)$ and the projected correlation function $\Xi(\sigma)$ of the 1,635 source sample of radio galaxies with spectroscopic redshifts from GAMA.}
\begin{tabular}{l c c c r} \hline
Function & Range & $s_0$/$r_0$ & $\gamma$ & $b(z=0.34)$ \\ 
&  ($h^{-1}$Mpc) &  ($h^{-1}$Mpc) & & \\ \hline
$\xi(s)$ & $2<s<300$ & $7.80^{+0.40}_{-0.53}$ & $1.34^{+0.05}_{-0.05}$ & $1.72^{+0.08}_{-0.08}$\vspace{5pt}\\
$\Xi(\sigma)$ & $\sigma<50$ & $8.16^{+0.40}_{-0.46}$ & $1.60^{+0.04}_{-0.05}$ & $1.92^{+0.10}_{-0.11}$\\ \hline
\label{xipar}
\end{tabular}
\end{center}
\end{table}

%
%

Redshift-space distortions affect our measurements of galaxy separations only in the direction of their lines of sight. This can be ameliorated by calculating the correlation function as a function of the line-of-sight separation, $\pi$, and the transverse separation, $\sigma$. Integrating over the $\pi$ coordinate gives the projected correlation function, which is a function of the ostensibly real-space $\sigma$ coordinate:

\begin{equation}
	\Xi(\sigma) = 2 \int^\infty_0 \xi(\sigma,\pi) \mathrm{d}\pi .
\end{equation}

For practical purposes we must impose our own upper limit on $\pi$ for the integration in order to strike a compromise between capturing the full clustering signal and avoiding the introduction of excessive noise at the highest separations. \citet{hawkins03}, for example, find their results insensitive to $\pi_\textrm{max} > 60$ $h^{-1}$Mpc. We use a limit of $\pi_\textrm{max} = 50$ $h^{-1}$Mpc in this work, avoiding considerable noise in $\xi(\pi>50,\sigma)$ . This function, free of redshift-space effects can be related to the real-space correlation function \citep{davis83}:

\begin{equation}
\frac{\Xi(\sigma)}{\sigma} = \frac{2}{\sigma} \int^\infty_0 \frac{r \xi(r )}{\sqrt{r^2 - \sigma^2}} \mathrm{d}r .
\end{equation}

Assuming a power law form for $\xi(r )$ as in Equation \ref{eq:xir}, we can fit $\Xi(\sigma)/\sigma$ to find the real-space correlation length, $r_0$, and power law slope, $\gamma$:

\begin{equation}
\frac{\Xi(\sigma)}{\sigma} = \left(\frac{r_0}{\sigma}\right)^\gamma H_\gamma .
\end{equation}

Naturally, the requirement of having spectroscopic redshifts to accurately determine the galaxy pair separations limits the use of this method to those 1,635 radio sources with GAMA spectroscopy. Furthermore, by binning galaxy pairs in two dimensions and then integrating, we sacrifice signal-to-noise in any given $\sigma$ bin, limiting the scope of the available data. We most likely underestimate the errors associated with $\Xi(\sigma)$, however, as the errors on the numerous $\pi$ bins being integrated are treated as independent. 

Given the advantages gained by having spectroscopic redshifts, we have measured the redshift-space and projected correlation functions of those radio sources with optical counterparts in the GAMA survey itself (Table \ref{xipar} and Figure \ref{fig:xis}). For this sample at $z\sim0.34$, we find a redshift-space correlation length $s_0 = 7.80^{+0.40}_{-0.53}$ $h^{-1}$Mpc with $\gamma \sim 1.34$, and correcting for redshift-space distortions, for the projected correlation function we find $r_0 = 8.16^{+0.40}_{-0.46}$ $h^{-1}$Mpc with $\gamma \sim 1.60$.

\subsection[]{Results and Discussion}
Our more direct spatial clustering measures at low redshift using the spectroscopic sample appear to agree with the results from the deprojection of the angular correlation function, $w(\theta)$, albeit with the latter constraints on $r_0$ and bias being much weaker. The results of this spectroscopic sample lend themselves to comparison with the work of \citet{brand05} and \citet{magliocchetti04}. \citet{brand05} directly measure the spatial correlation function (although ignoring redshift-space distortions) of 268 radio galaxies in 165 deg$^2$ of the Texas-Oxford NVSS Structure (TONS) survey with flux density $S_{1.4} > 3$ mJy and an optical limit of $R \lesssim 19.5$. The slightly higher radio flux limit removes some of the sources associated with star formation, but the brighter optical limits favour low-luminosity radio galaxies at lower redshift, with the median of the sample at $z\sim0.3$. They find $r_0 = 6.1 \pm 1.1$ $h^{-1}$Mpc and assume linear ($\epsilon = \gamma - 1$) clustering model in good agreement with our corresponding measure with the ``FIRST--GAMA--matched--z$<$0.5'' sample. 

\citet{magliocchetti04} make joint use of FIRST and 2dFGRS to a radio flux density limit of 1 mJy and an optical limit of $b_J < 19.37$, working with a sample of 820 radio sources with redshifts $0.01 < z < 0.3$ over a larger area of $\sim 375$ deg$^2$. Their value of $r_0 \sim 4.7 \pm 0.7$ $h^{-1}$Mpc is lower than for our sample, but perhaps by no more than expected given their lower redshifts and preferential selection of optically brighter sources. This effect is mitigated by their subsequent selection of only those sources whose spectra have signatures of AGN activity. This increases the redshift from $z\sim0.10$ to $0.13$, and the correlation length to $r_0 \sim 7.6 \pm 0.8$ $h^{-1}$Mpc, similar to that of our sample $z\sim0.34$, shown in Figure \ref{r0}. We would expect to observe clustering between these two values, albeit boosted due to our sample having a higher median redshift. The close agreement with these past surveys also suggests that our analysis accounting for the noise variations in the GAMA fields is robust.

\section[]{Mass Bias}\label{massbias}

	The differences in clustering of different classes of extragalactic objects and the background matter distribution motivates the use of some bias parameter, $b$, as introduced by \citet{kaiser84} and \citet{bardeen86}: 
	\begin{equation}
		b^2(z) = \frac{\xi_{\textrm{gal}}(r,z)}{\xi_{\textrm{DM}}(r,z)}, \label{eq:bias}
	\end{equation}
	where the numerator and denominators are the galaxy and dark matter correlation functions, respectively. 
	
	The bias parameter can be determined directly if both the real- and redshift-space forms of the correlation function are known. The ratio of the two can be approximated by a quadratic in the infall parameter $\beta \propto b(z)$ \citep{kaiser87}.
	Therefore, with a good estimate of the real- and redshift-space correlation functions, one may calculate the ratio of the two and thereby the bias. 
	
We are able to infer the real-space correlation length $r_0$ from the angular correlation function (Section \ref{Limber}) or the projected correlation function (Section \ref{projxi}). The bias parameter (as a function of redshift) may be defined as Equation \ref{eq:bias} with $r=8$ $h^{-1}$Mpc. As described in Equation \ref{eq:xir} the numerator can be written
	\begin{equation}
	 \xi_{\textrm{gal}} (8,z) = \left[ \frac{r_0 (z)}{8} \right]^\gamma .
	 \end{equation}
	 The corresponding function for the denominator is given by \citet{peebles80} as
	 \begin{equation}
	  \xi_{\textrm{DM}} (8,z) = \sigma_8^2 (z) / J_2
	  \end{equation}
	  where $J_2 = 72/[(3-\gamma)(4-\gamma)(6-\gamma)2^\gamma]$ and the parameter $\sigma_8^2$ is the dark matter density variance in a comoving sphere of radius 8 $h^{-1}$Mpc. 
	  The combination of these equations gives the scale-independent evolution of bias with redshift, given only the correlation length and slope:
	  \begin{equation}
	  	b(z) = \left[ \frac{r_0(z)}{8} \right]^{\gamma / 2} \frac{J_2^{1/2}}{\sigma_8 D(z) / D(0)}.
	\end{equation}
	In our subsequent analysis, the redshift value is assumed to be the median of the distribution of objects.
	
	\begin{figure*}
\begin{minipage}[b]{0.48\linewidth}
\centering
\includegraphics[width=\textwidth]{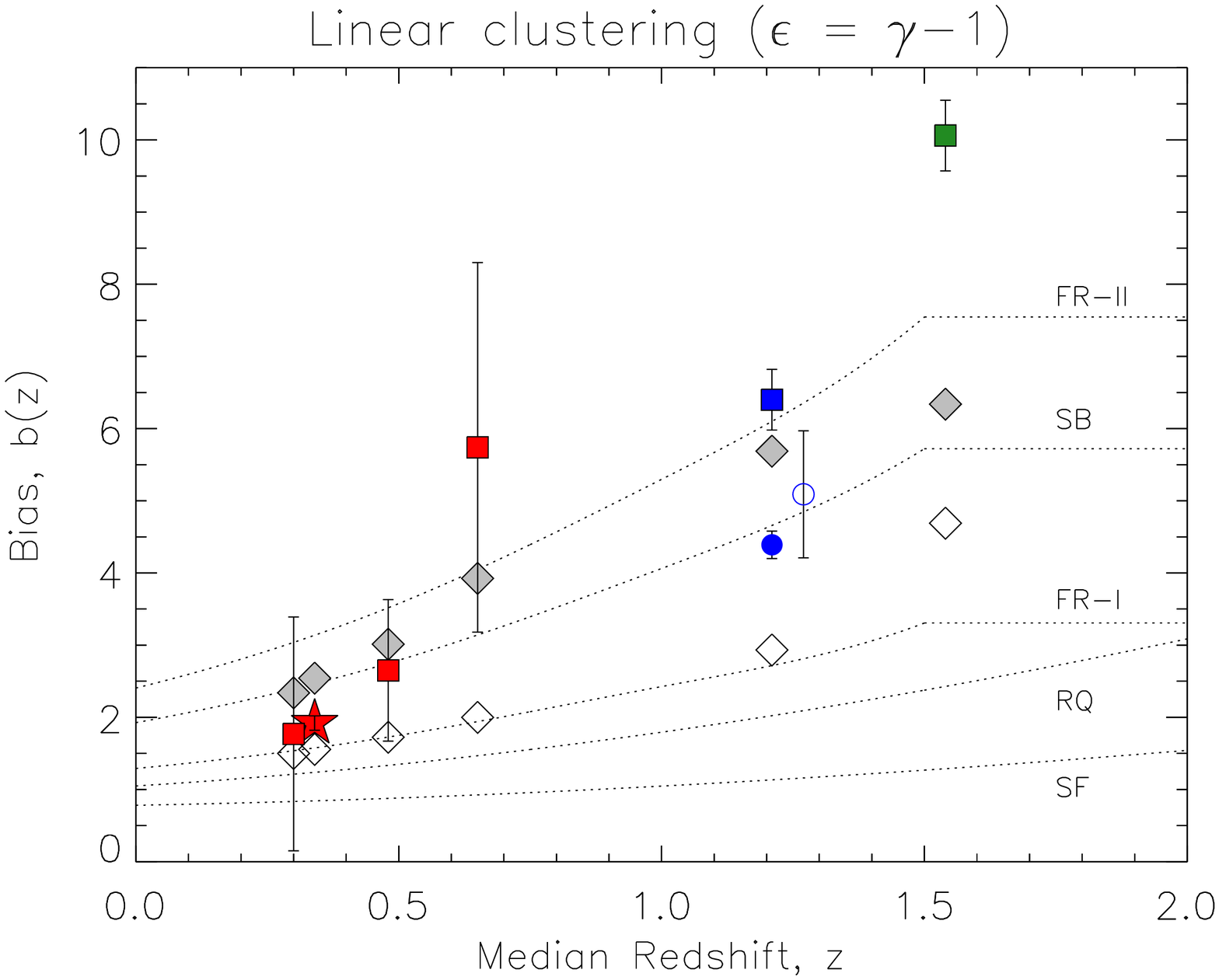}
\end{minipage}
\begin{minipage}[b]{0.48\linewidth}
\centering
\includegraphics[width=\textwidth]{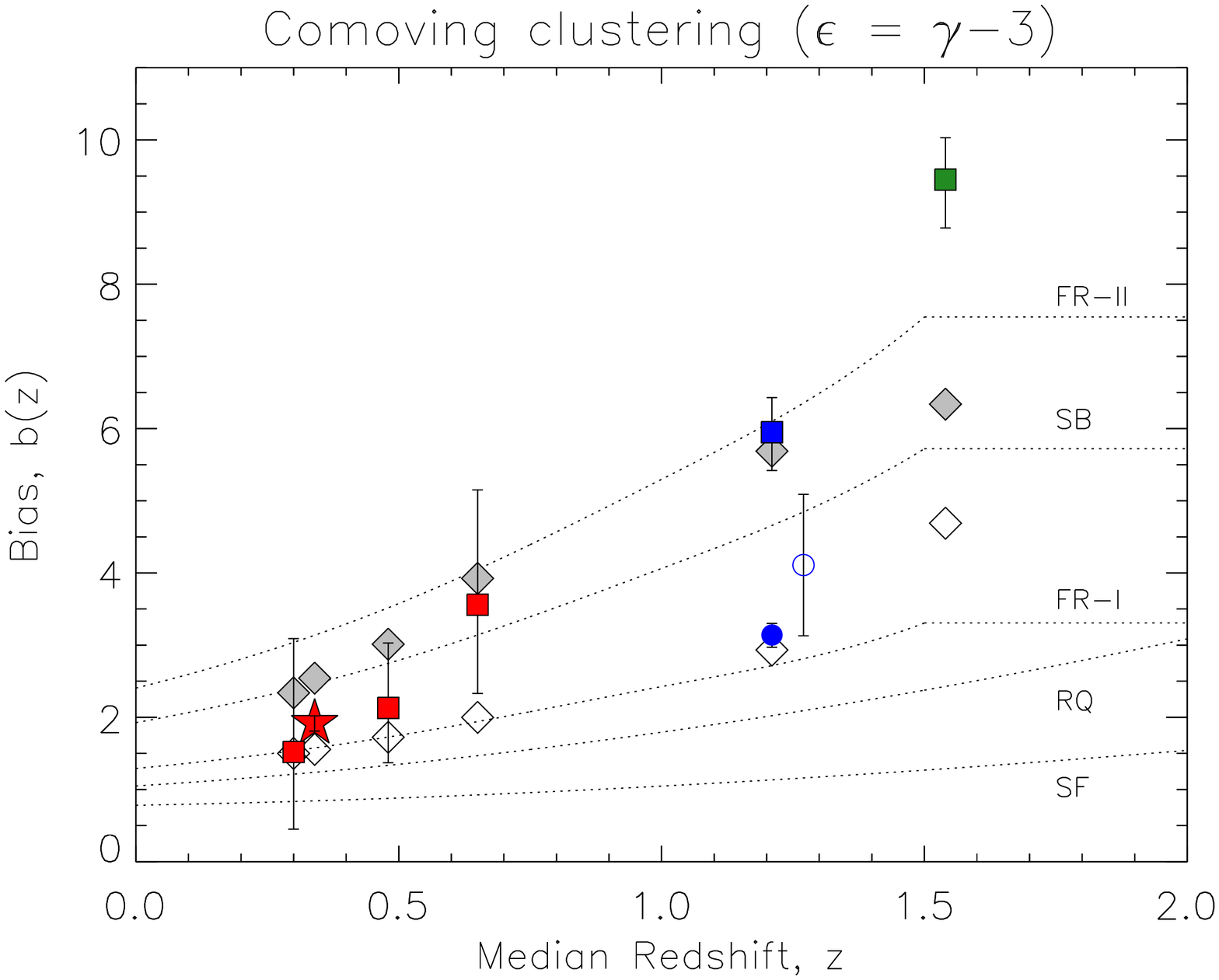}
\end{minipage}
  \caption{Linear bias parameter for the observed radio samples described in Table \ref{parameters} as calculated for two different clustering indices, with symbols as defined in Figure \ref{r0}. Overplotted are lines showing the model bias evolution based on the halo masses assigned in the SKADS simulation of individual source populations: FR{\sc I}, FR{\sc II} and radio-quiet (RQ) AGN, starbursts (SB) and normal star-forming (SF) galaxies. Open diamonds corresponding to each point mark the aggregate bias expected of the samples, assuming relative population abundances and masses used in redshift-matched SKADS samples. The grey filled diamonds show this same model bias if we increase the assumed halo mass of FR\textsc{I} sources to equal that of the FR\textsc{II} sources.}
  	\label{bias}
\end{figure*}
	
\subsection{Results}

The bias inferred from the angular correlation function is shown in Table \ref{parameters} for both $\epsilon$ values corresponding to linear and comoving clustering evolution. These results are compared in Figure \ref{bias} with the bias model by \citet{mo96} used by \citet{wilman08} in populating dark matter haloes in the SKADS simulations. They assign a particular halo mass to each source type in the simulation: FR\textsc{I}, FR\textsc{II} and radio-quiet AGN, normal star-forming galaxies and starbursts, with assumed halo masses of $10^{13}$, $10^{14}$, $3\times 10^{12}$, $10^{11}$ and $5 \times 10^{13}$ $h^{-1}M_\odot$, respectively, and impose a plateau for each model at high redshifts above which the assumption of a fixed halo mass breaks down. The model $b(z)$ for each of these source types is denoted by a black dotted line, showing the stronger bias for those objects residing in more massive haloes.  

By weighting the SKADS sources by the ratio between the sample and SKADS redshift distributions, we have estimated the relative proportions of the population masses. Calculating the model bias of the sample as the weighted mean of the biases of each individual population, we show a predicted bias at the 5 different redshifts probed in Figure \ref{bias} (\textit{open diamonds}), accounting for the change in galaxy halo mass being observed. These follow the same trend as our measurements, with the bias rising more steeply with redshift than any individual model population owing to the increasing AGN fraction towards higher $z$. Our measurements over the GAMA fields agree qualitatively with the model, but with those associated only with the GAMA fields exceeding the predictions to varying degrees. For those sources with measured redshifts, the difference is within the stated errors, but at higher redshifts our measurements exceed the SKADS bias values by $\sim 3\sigma$ (we return to this point in Section \ref{biasdiscussion}).

	
The spatial correlation function of spectroscopic sources provides one additional data point for the bias at a redshift of $z=0.34$, similar to the $z<0.5$ subsample of combined spectroscopic and photometric matches. Table \ref{xipar} shows $r_0$ and $s_0$ derived from $\xi(s)$ and $\Xi(\sigma)$ respectively, and the inferred bias (where redshift-space effects are wrongly ignored in the case of the former). We find a bias of $\sim1.7$ for the redshift-space calculation of the bias, with a real-space equivalent of $\sim1.9$ at $z\sim0.34$. The results from $\Xi(\sigma)$ are shown in Figure \ref{bias} for comparison with the angular clustering results.

\subsection{Discussion}\label{biasdiscussion}

Within our matched sample and incorporating simulated data, we observe the increasing of $b(z)$ in line with models corresponding to fixed masses \citep{mo96, matarrese97}, although we find a relatively  high value for the high-redshift sources, particularly when using the linear clustering parameter. Whilst the GAMA data points do not place especially tight constraints on this evolution, and the well-constrained point from $\Xi(\sigma)$ is limited to one epoch, the assumption of the redshift distributions for the full and unmatched subsamples probe redshifts up to $z = 1.55$ and give more shape to the evolution. The agreement is good at lower redshifts, the GAMA-matched results (\textit{red}) lying $<1\sigma$ from the model prediction as well as agreeing with the 2SLAQ radio-detected LRGs examined by \citet{wake08} ($b\sim3.0$). The higher redshift points, however, significantly exceed the values prescribed in the SKADS simulation. The wider FIRST sample is an exception, falling significantly below the trend displayed by the narrower samples, but closely matching the Mo \& White bias prescription from SKADS. On face value, this apparent excess clustering suggests a greater proportion of AGN observed than was assumed in the simulations, or perhaps a considerable underestimate of the typical halo masses of any or all of the galaxy types considered.

Our unmatched FIRST sources, especially, in the GAMA field are more strongly clustered than would be expected by simply subtracting the cross-matched $N(z)$ from the SKADS $N(z)$ and assuming the same population as a matching SKADS sample. This means either that the assumed redshift distribution is skewed towards low redshift, or the fraction of more massive galaxies is higher than expected. Inspection of Figure~\ref{zdists} shows that our assumed $N(z)$ for this point contains a significant fraction of $z<0.5$ sources where one would expect to observe a greater proportion of less massive, star-forming galaxies rather than the more massive AGN. Given that if these low-redshift objects were detected in FIRST, we would expect them to have optical counterparts in SDSS/GAMA (cf. almost complete cross-identification at $z\sim0.6$ in Figure \ref{zdists}), it is perhaps unrealistic to assume these missing low-$z$ sources remain in our unmatched sample. The cause of this large fraction of unmatched sources is discussed in Section~\ref{redshifts}, however to determine the effect that these sources have on our measured clustering length and bias we implement cuts on the SKADS redshift distribution, where we remove all sources at $z< 0.05$ and $z< 0.1$ from the simulations. This is a coarse way of simulating the combined effect of the resolution bias and the removal of very low-redshift sources with photometric redshifts with $z<0.002$. We then recalculate the correlation length and bias using these new redshift distributions and find that the correlation length and bias increases significantly ($\sim$25--50 per cent increase in bias). 
Although subject to many uncertainties, this may point to a higher proportion of highly biased objects in the radio source population at high redshift. 

One of the main uncertainties in the SKADS simulation is indeed the evolution and bias of the FR\textsc{I}-type objects which dominate the source counts at the flux density limit of the FIRST survey at $z>0.5$. In the SKADS simulation the FR\textsc{I}s are less biased than the FR\textsc{II}s, however it is becoming clear that there is a large overlap in the stellar mass distributions between the generally less radio luminous FR\textsc{I}s and their FR\textsc{II} counterparts at high redshifts \citep[e.g.][]{mclure04,herbert11}, thus it is possible that the bias prescription for the FR\textsc{I} sources in the SKADS simulation is underestimated. If we assign a similar halo mass to the FR\textsc{I} as for the FR\textsc{II}s then we find that the expected average bias of radio sources in our sample ($z\sim1.21$) to be in the region of $b=5.7$, this is much more closely aligned with our measured value of $b=5.95^{+0.48}_{-0.53}$. 

Deeper surveys over smaller areas will be able to address this issue better than the relatively shallow FIRST data. Indeed, deep multi-wavelength surveys are now beginning to tackle the question of redshift evolution \citep[e.g.][]{smolcic09,mcalpine11,simpson12,mcalpine13} and with slightly more area could measure the clustering length.


\section[]{Conclusions}\label{conclusions}

Using radio observations from the FIRST survey and optical/infrared data (SDSS/UKIDSS LAS) over the GAMA survey field area, we have obtained a cross-matched sample of radio galaxies with optical host galaxy redshifts ($\sim$42 per cent spectroscopic). We have measured the redshift-space and projected correlation functions of these spectroscopic identifications from GAMA, as well as the angular correlation function of radio sources over the extended GAMA survey area. Assuming parent redshift distributions from SKADS \citep{wilman08}, we have inferred the spatial correlation length $r_0$ and the mass bias $b(z)$ for the matched radio sources and the radio sources without optical identifications, extending our redshift range up to $z \sim 1.6$. The results can be summarized as follows:
\begin{enumerate}
	\item The projected correlation function of FIRST sources with GAMA spectroscopic counterparts provides a well-constrained correlation length and linear bias of $r_0 = 8.16^{+0.40}_{-0.46}$ $h^{-1}$Mpc and $b = 1.92^{+0.10}_{-0.11}$ at $z \sim 0.34$.
	\item The angular correlation function follows a power law but we find a steeper slope of $\sim-1.2$ compared with that found for other classes of galaxy and often assumed in the literature.
	\item Our cross-matched sample yields a spatial correlation length of $r_0 \sim 8.5$ $h^{-1}$Mpc to $r_0 \sim 10.7$ $h^{-1}$Mpc at $0.3 <z< 0.65$ (assuming stable clustering evolution). Adding the assumption of the SKADS parent $N(z)$, we measure the clustering length up to median redshift $z \sim 1.55$, where we find $r_0 \sim 12.2$ $h^{-1}$Mpc. 
	\item We measure the bias as a function of redshift across the subsamples finding it to increase from $b(z=0.30) \sim 2.8$ to $b(z=1.55) \sim 9.2$. These values were compared with predicted values from a model assuming population fractions and masses from SKADS and were in qualitative agreement but exceeding the prescribed values at high redshift. This is most probably due to a combination of surface density fluctuation in the FIRST survey, leading to a shortfall of sources at $\sim 1$mJy which in turn biases our sample towards a higher fraction of strongly clustered AGN at high redshifts and/or potential inaccuracies in the halo masses of particular radio subpopulations used in SKADS.
          \item If we assign a similar halo mass to the FR\textsc{I} sources as assumed for the FR\textsc{II} sources in our radio survey then we find that we can reproduce our bias value at high redshift. A more highly biased FR\textsc{I} population may in turn lead to a highly biased tracer of the high-redshift Universe for cosmological applications with radio surveys \citep[e.g.][]{raccanelli12, camera12}.
\end{enumerate}

While we place modest constraints on the clustering evolution of mJy radio sources and suffer from limited cross-identification with the optical surveys, our use of a well-constrained redshift distribution from SKADS allows us to extend our redshift range to $z > 1$. This highlights some potentially important discrepancies with a simple model bias, implying stronger clustering at higher redshifts than is expected from the fixed halo masses assumed in the SKADS simulations through some combination of a greater proportion of massive AGN than previously thought and/or an increased typical halo mass being observed. In a future paper, we analyse deep field data to help confirm our high-redshift bias estimates.

\section*{Acknowledgements}

GAMA is a joint European-Australasian project based around a spectroscopic campaign using the Anglo-Australian Telescope. The GAMA input catalogue is based on data taken from the Sloan Digital Sky Survey and the UKIRT Infrared Deep Sky Survey. Complementary imaging of the GAMA regions is being obtained by a number of independent survey programs including GALEX MIS, VST KIDS, VISTA VIKING, WISE, Herschel-ATLAS, GMRT and ASKAP providing UV to radio coverage. GAMA is funded by the STFC (UK), the ARC (Australia), the AAO, and the participating institutions. The GAMA website is http://www.gama-survey.org/ .

\balance
\footnotesize{

\bibliographystyle{mn2e}
\bibliography{../../../biblio}	
}
	
\label{lastpage}

\end{document}